\documentclass[12pt,fleqn,leqno]{article}
\usepackage{amsmath,amssymb}
\usepackage{amsthm}
\usepackage[dvips]{graphicx}

\allowdisplaybreaks

\numberwithin{equation}{section}

\makeatletter
\renewcommand{\section}{\@startsection{section}{1}{0pt}{20pt}{6pt}{\large\bf}}
\renewcommand{\@seccntformat}[1]{\csname the#1\endcsname.\ }

\def\footnoterule{\kern -3pt \hrule width 2.7 true cm \kern 2.6pt}

\def\ni{\noindent}
\def\vs{\vspace}
\def\hs{\hspace}

\def\EE{\mathsf E}
\def\PP{\mathsf P}
\def\QQ{\mathsf Q}

\def\cE{\mathcal{E}}
\def\cC{\mathcal{C}}
\def\R{I\!\!R}
\def\L{I\!\!L}
\def\wt{\widetilde}
\def\wh{\widehat}

\newcommand{\eps}{\varepsilon}
\newcommand{\p}{\! +\! }
\newcommand{\m}{\! -\! }

\newtheorem{theorem}{Theorem}[section]

\newtheorem{example}[theorem]{Example}

\newtheorem{remark}[theorem]{Remark}

\begin{document}

\title{\textbf{On American VIX options under the generalized 3/2 and 1/2 models}}
\author{J\'er\^ome Detemple\thanks{Questrom School of Business, Boston University, Boston, USA. E-mail: \texttt{detemple@bu.edu}}\quad \& Yerkin Kitapbayev\thanks{Questrom School of Business, Boston University, Boston, USA. E-mail: \texttt{yerkin@bu.edu}}}
\maketitle

\vs{-18pt}
\begin{center}
\textbf{Abstract}
\end{center}

{\par \leftskip=2.6cm \rightskip=2.6cm \footnotesize
In this paper, we extend the 3/2-model for VIX studied by Goard and Mazur (2013) and introduce the
generalized 3/2 and 1/2 classes of volatility processes. Under these models, we study the pricing of European and American VIX options and,
for the latter, we obtain an early exercise premium representation using a free-boundary approach and local time-space calculus. The optimal exercise boundary
for the volatility is obtained as the unique solution to an integral equation of Volterra type.

We also consider a model mixing these two classes
and formulate the corresponding optimal stopping problem in terms of the observed factor process. The price of an American VIX call is then represented
by an early exercise premium formula. We show the existence of a pair of optimal exercise boundaries for the factor process and characterize them as the unique solution
to a system of integral equations.

\par}


\vs{6pt}




\vs{-10pt}

\section{Introduction}

During recent decades, financial markets have experienced significant fluctuations in volatility.
These events have spurred demands for volatility indicators and for derivative instruments to manage volatility risk.
Nowadays, the most popular volatility measurement is the VIX, which is the implied
volatility of 30-day S\&P500 options. VIX futures contracts started to trade on March 26, 2004 on the CBOE.
Options on the VIX, introduced on February 24, 2006, also by the CBOE, have proven
increasingly popular with investors. Since their introduction, volume has
grown from a daily average of 23,491 contracts in 2006 to 632,419 in 2014.
This popularity stems in part from the recurrence of rapidly changing
volatility episodes, especially during the recent crisis. VIX options
provide an effective way to manage risks tied to volatility fluctuations.

The valuation of VIX options has been considered well before their actual
introduction on the CBOE. The issue became of interest in the early 90s,
around the time when the VIX index was introduced to measure volatility (see
Whaley (1993)). Valuation formulas have developed around a set of well known
models for the evolution of the underlying volatility. Formulas for European
volatility options can be found in Whaley (1993) under the assumption of a
geometric Brownian motion process (GBMP) and in Gr\"{u}nbichler and Longstaff
(1996) for a mean-reverting square-root volatility process (MRSRP), also
known as CIR process\ (see Cox, Ingersoll and Ross (1985)). For
American-style volatility contracts, Detemple and Osakwe (2000) provide
formulas for Geometric Brownian motion process (GBMP), mean-reverting
Gaussian process (MRGP), mean-reverting square root process (MRSRP) and
mean-reverting log process (MRLP). All these cases can be embedded in the
volatility models,%
\begin{equation}\hs{6pc}
dX_{t}=\left( \beta -\alpha X_{t}\right) dt-\kappa X_{t}^{\gamma }dB_{t},\ \
\ \ \ \text{with }\gamma =0,1/2\text{ or }1  \label{gam}
\end{equation}%
\begin{equation*}\hs{6pc}
d\ln X_{t}=\left( \beta -\alpha \ln X_{t}\right) dt-\kappa dB_{t}.
\end{equation*}%
This last one is an exponential transform of a Gaussian process. Another
transform of volatility that has been used to price variance contracts is
the Heston (1993) model, where the local variance $v=X^{2}$ follows,%
\begin{equation*}\hs{6pc}
dv_{t}=\left( \beta -\alpha v_{t}\right) dt-\kappa v_{t}^{1/2}dB_{t}
\end{equation*}%
which is a MRSRP for $v$. Contracts on realized variance or realized
volatility, such as swaps and options, have been examined under this
specification by several authors, including Broadie and Jain (2008).
Realized variance/volatility contracts have also been priced under various
generalizations of the Heston model, e.g., Elliott, Siu and Chan (2007) and
Sepp (2008).

More recently, Goard and Mazur (2013) examine the valuation of VIX options
under the $3/2$ specification,%
\begin{equation}\label{3/2}\hs{6pc}
dX_{t}=\left( \alpha X_{t}-\left( \beta \m\kappa ^{2}\right) X_{t}^{2}\right)
dt+\kappa X_{t}^{3/2}dB_{t}.
\end{equation}%
This process was originally introduced by Heston (1997) and Platen (1997) to
model the evolution of the local variance $v$ of an asset return. An
interesting feature of the process is that it allows for spikes, a property
found in volatility data. Drimus (2001) provides empirical evidence in favor
of the model for FX markets. Goard and Mazur, show that the 3/2 specification, as a
model for volatility, provides a better fit to the VIX data than various
alternatives including GBMP, MRGP, MRSRP, MRLP and Heston's MRSRP for $v$.
They also compute European VIX option prices under this model.
Relying on this evidence, Liu (2015) formulates a free-boundary problem for the valuation of the American VIX
put option under the 3/2 model and shows monotonicity properties of the option price function and optimal
exercise boundary.

Although the evidence provided in Goard and Mazur (2013) shows that the 3/2
model dominates the alternatives considered, the analysis performed tests
for overidentifying restrictions relative to a specific benchmark. This
benchmark has a more general structure that nests the various alternatives
tested. It nevertheless imposes specific functional forms on the
coefficients of the VIX process. Unconstrained GMM shows that the benchmark
has an estimated $\gamma $ of 1.48 and places large weights on the various
nonlinear components in the drift. These results suggest that specifications
deviating from the standard 3/2 model are of interest for capturing complex
aspects of the VIX behavior.

A recent step in that direction is taken by Grasselli (2015), who introduces the 4/2 model for the local variance process, which is the sum of a 1/2 and a 3/2 models. Instantaneous volatility, in the 4/2 model, is $a\sqrt{Y} + b/\sqrt{Y}$ where $a, b$ are positive constants and $Y$ follows a CIR process. This model has several interesting features. Most notably, variance is bounded away from zero, as suggested by the stylized facts reported in Gatheral (2008). The model also helps to explain observed shapes of the implied volatility surface. Grasselli (2015) studies the behavior of the 4/2 price process and derives the characteristic function of the log price. He also provides an exact simulation scheme based on the conditional distribution of the price.

This paper has several contributions. First, it introduces two new
classes of volatility processes, the generalized 3/2 class $(A1)$
and the generalized 1/2 class $(A2)$. These two classes contain a variety of
processes that are natural extensions of the 3/2 and 1/2 processes, yet
remain tractable for valuation purposes. The computations of vanilla options and futures on VIX can be executed efficiently by standard numerical integration methods. Also,
we note that models in the generalized 3/2 class produce a positive skew of implied volatilities which is the most relevant stylized fact of the VIX market, as documented by Mencia and Sentana (2013).
Second, it provides explicit
formulas (in the form of integrals) for European and American call and put options when the underlying volatility
follows any process in $(A1)$ or $(A2)$. In the American case, an early exercise
premium representation formula is derived using the free-boundary approach and local time-space calculus (see Peskir (2005a)). The optimal exercise boundary for the VIX process is characterized
as the unique solution to a nonlinear integral equation of Volterra type. Third, we show that the value function of the optimal stopping problem
satisfies a smooth-fit property along the optimal exercise boundary in the case where the dependence on the  initial value of the underlying process
is unknown. To the best of our knowledge, existing papers considered problems where the underlying processes have
explicit initial dependence, e.g., Brownian motion, geometric Brownian motion, etc. Another aspect, outlined in the paper, is that numerical computations show a non-convexity of the American call price function with respect to the initial value of the VIX under the 3/2 model (see Figure 3),
but not in the 1/2 model (see Figure 4).

The final section of the paper is devoted to the pricing of the American call when the VIX is modelled as the mixture of the two classes of models above, i.e., the sum of generalized 3/2- and 1/2-type processes. Equivalently, the VIX process is a function of a CIR process where this function is the sum of functions of $(A1)$ and $(A2)$ types. We show that, under certain assumptions, there exists a pair of optimal exercise boundaries for the underlying CIR process that can be obtained as the unique solution to a system of coupled integral equations.  We then provide the early exercise premium representation formula for the American call price. This formula decomposes it into the sum of a European part and an early exercise premium which depends on the pair of exercise boundaries.

The paper is organized as follows. Section 2 describes the two classes of
processes that are the focus of this study, formulates the pricing
problem for an American VIX call as an optimal stopping problem and shows how to price a European VIX call. An
associated free-boundary problem for the American call option is studied in Section 3. Section 4 derives
the early exercise premium representation for the American call price and
characterizes the optimal exercise boundary as the unique solution to a
nonlinear integral equation. Section 5 provides corresponding results for
European and American put options.  Section 6 studies the VIX call price under the mixture model. The paper is completed by a technical appendix.
\vs{6pt}

\section{The generalized 3/2 and 1/2 models and VIX options}

1. First let us consider the following two classes of functions
\vs{4pt}

$(A1)$ Generalized 3/2-type: let $f\left( \cdot \right) :\R_{+}\rightarrow \R_{+}$ be a
three times continuously differentiable, strictly decreasing and convex function ranging
from $+\infty $ to $0$. Let $g\left( \cdot \right) $ be the inverse of $%
f\left( \cdot \right) $, i.e. $f\left( g\left( x\right) \right) =x$ for $x>0$.
\vs{4pt}

$(A2)$ Generalized 1/2-type: let $f\left( \cdot \right) :\mathbb{R}_{+}\rightarrow \mathbb{R}_{+}$ be a
three times continuously differentiable, strictly increasing and weakly concave function
ranging from $0$ to $\infty $. Let $g\left( \cdot \right) $ be the inverse
of $f\left( \cdot \right) $, i.e. $f\left( g\left( x\right) \right) =x$.
\vs{4pt}

In this paper we model the VIX, under the historical measure $\PP$, as follows
\begin{align}\label{vix}\hs{6pc}
X_{t} =f(Y_t)
\end{align}
for $t\ge 0$ where $f$ is either of type $(A1)$ or $(A2)$ above and
a factor process $Y=(Y_t)_{t\ge 0}$ is given by
\begin{equation} \label{CIR} \hs{6pc}
 dY_{t}=\big(\beta-\alpha Y_t\big)dt-\kappa\sqrt{Y_t} dB_{t},\quad Y_0=y
 \end{equation}
where $\alpha,\beta ,\kappa>0 $ are constant parameters and $B$ is a $\PP$- standard Brownian motion (SBM). The process $Y$ solving \eqref{CIR} follows a mean-reverting square-root process (MRSRP) and the random variable $Y^y_t$ has non-central chi-squared density function
 $q(\wt{y};t,y)$ (see, e.g., Cox, Ross and Ingersoll (1985)). Throughout this paper, we assume that $\beta\ge \kappa^2/2$ as Feller showed that under this condition $Y$ is strictly positive. Hence $X$ is well defined and is strictly positive for all $t>0$.  We note that the functions $f$ are strictly monotone so that there is a one-to-one relationship between
the VIX process $X$ and the factor process $Y$.

By using Ito's formula, we get the dynamics of $X$
\begin{align}\label{vix-1}\hs{4pc}
dX_{t} =&\left( f^{\prime }\left( g\left( X_{t}\right) \right) \left( \beta
\m\alpha g\left( X_{t}\right) \right) +\frac{1}{2}\kappa ^{2}f^{\prime \prime
}\left( g\left( X_{t}\right) \right) g\left( X_{t}\right) \right) dt
\\
&\;-\kappa f^{\prime }\left( g\left( X_{t}\right) \right) \sqrt{g\left(
X_{t}\right) }dB_{t}\nonumber
\end{align}%
for $t\ge 0$.
As $X$ is not the price of a traded asset, one should allow for the possibility of a non-zero market price of risk $\lambda(t,X)$ associated with the VIX. Following papers by Stein and Stein (1991) and Gr\"{u}nbichler and Longstaff (1996), we assume that the market price of risk is such that the risk-neutral process for $X$ is of the same form as the real process \eqref{vix-1}. For this, one chooses $\lambda(t,X_t)$ as
$a/\sqrt{g(X_t)}+b\sqrt{g(X_t)}+c f^{\prime \prime}( g( X_{t}))\sqrt{g( X_t)}/ f^{\prime}( g( X_{t}))$. We recall that Egloff, Leippold and Wu (2010) and Mencia and Sentana (2013)
showed evidence that the price of risk related to the VIX is negative. It is clear from our specification that the negative sign can be easily obtained. To avoid additional notations, we assume that the dynamics \eqref{vix-1} is under some risk neutral measure $\QQ$ and $B$ is $\QQ$-SBM.

The specification \eqref{vix-1} of type $(A1)$  includes several models of potential interest
to describe the evolution of the VIX.

\begin{example}
($3/2$-model) The 3/2-model is introduced by Goard and Mazur (2013). It is
obtained by taking $f\left( y\right) =1/y$. Then, $g\left( x\right) =1/x$, $%
f^{\prime }\left( y\right) =-1/y^{2}$, $f^{\prime \prime }\left( y\right)
=2/y^{3}$ and
\begin{align}\label{ex-1}\hs{4pc}
dX_{t}=\left( \alpha X_{t}-\left( \beta \m\kappa ^{2}\right) X_{t}^{2}\right)
dt+\kappa X_{t}^{3/2}dB_{t}.
\end{align}%
The 3/2 model has elasticity of variance equal to $\varepsilon =3$. It also
displays mean reversion if $\beta >\kappa ^{2}$. The speed of mean reversion
$\left( \beta -\kappa ^{2}\right) X_{t}$ is linear in the VIX. The constant
attractor is $\alpha /\left( \beta \m\kappa ^{2}\right) $.
\end{example}

\begin{example}
($1 \p1/(2\nu)$-model) Let $f\left( y\right) =1/y^{\nu }$ where $\nu >0$ and
$\beta >\frac{1}{2}\kappa ^{2}\left( \nu +1\right) $. Then, $g\left(
x\right) =\left( 1/x\right) ^{1/\nu }$, $f^{\prime }\left( y\right) =-\nu
/y^{\nu +1}$, $f^{\prime \prime }\left( y\right) =\nu \left( \nu \p1\right)
/y^{\nu +2}$ and%
\begin{align}\label{ex-2}\hs{4pc}
dX_{t}=\nu \left( \alpha X_{t}-\left( \beta -\frac{1}{2}\kappa ^{2}\left(
\nu +1\right) \right) X_{t}^{1+1/\nu }\right) dt+\nu \kappa X_{t}^{1
+1/(2\nu)}dB_{t}.
\end{align}%
For this specification the elasticity of variance is $\varepsilon =2\p1/\nu$.
The process has linear speed of mean reversion $\nu \left( \beta \m\frac{1}{2}%
\kappa ^{2}\left( \nu \p1\right) \right) X_{t}$ and constant attractor $%
\left(\alpha /\left( \beta \m\frac{1}{2}\kappa ^{2}\left( \nu \p1\right) \right)\right)^\nu $.
The $3/2$ model is obtained when $\nu =1$.
\end{example}

\begin{example}
(mixture $1+1/(2\nu _{j})$, $j=1,...,n$ model) Let $f\left( y\right)
=\sum_{j}\omega _{j}/y^{\nu _{j}}$ where $\nu _{j}>0$, $\omega
_{j}>0,j=1,...,n$ so that
\begin{align*}\hs{2pc}
\sum_{j}\omega _{j}\frac{1}{g\left( x\right) ^{\nu _{j}}}=x,\ \ \ \ \
f^{\prime }\left( y\right) =-\sum_{j}\omega _{j}\frac{\nu _{j}}{y^{\nu
_{j}+1}},\ \ \ \ \ f^{\prime \prime }\left( y\right) =\sum_{j}\omega _{j}%
\frac{\nu _{j}\left( \nu _{j}+1\right) }{y^{\nu _{j}+2}}
\end{align*}%
and%
\begin{align}\label{ex-3}\hs{-1pc}
dX_{t} =\left( \alpha \sum_{j}\omega _{j}\frac{\nu _{j}}{g\left(
X_{t}\right) ^{\nu _{j}}}-
\sum_{j}\omega _{j}\frac{\nu _{j}\left( \beta -\tfrac{1}{2}\kappa ^{2}\left( \nu _{j}+1\right)\right) }{g\left(
X_{t}\right) ^{\nu _{j}+1}}\right) dt
+\kappa \sum_{j}\omega _{j}\frac{\nu _{j}}{g\left( X_{t}\right) ^{\nu
_{j}+1/2}}dB_{t}.
\end{align}%
The elasticity of variance
is a non-linear function of the VIX. The process has non-linear speed of
mean reversion and a constant attractor.
\end{example}

The specification \eqref{vix-1} of type $(A2)$  contains another set of relevant models for the VIX.

\begin{example}
($1/2$-model) See, e.g., Grunblicher and Longstaff (1996). It is obtained by taking $f\left(
y\right) =y$, a weakly concave function. Then $g\left( x\right) =x$, $%
f^{\prime }\left( y\right) =1$, $f^{\prime \prime }\left( y\right) =0$ and%
\begin{align}\label{ex-4}\hs{5pc}
dX_{t}=\left( \beta -\alpha X_{t}\right) dt-\kappa X_{t}^{1/2}dB_{t}.
\end{align}
The $1/2$-model has elasticity of variance equal to $\varepsilon =1$. It also
displays mean reversion. The speed of mean reversion $\alpha $ is constant.
The long run mean is $\beta /\alpha $.
\end{example}

\begin{example}
($1\m 1/\left( 2\nu \right)$- model) Let $f\left( y\right) =y^{\nu }$ where $%
\nu \in \left( 0,1\right] $ and $\beta +\frac{1}{2}\kappa ^{2}\left( \nu
-1\right) >0$. Then, $g\left( x\right) =x^{1/\nu }$, $f^{\prime }\left(
y\right) =\nu y^{\nu -1}$, $f^{\prime \prime }\left( y\right) =\nu \left(
\nu \m1\right) y^{\nu -2}$ and
\begin{align}\label{ex-5}\hs{3pc}
dX_{t}=\nu X_{t}^{1-1/\nu }\left( \beta +\frac{1}{2}\kappa ^{2}\left( \nu
-1\right) -\alpha X_{t}^{1/\nu }\right) dt-\kappa \nu X_{t}^{1-1/\left( 2\nu
\right) }dB_{t}.
\end{align}
This model has non-linear elasticity of variance $%
\varepsilon =2\m 1/\nu$. It also displays
non-linear mean reversion with speed of mean reversion $\alpha \nu
X_{t}^{1-1/\nu }$. The attracting value is $\left( \left( \beta +\frac{1}{2}%
\kappa ^{2}\left( \nu -1\right) \right) /\alpha \right) ^{\nu }$. The $1/2$-
model is obtained for $\nu =1$.
\end{example}

\begin{example}
(mixture $1\m1/\left( 2\nu _{j}\right) $, $j=1,...,n$ model) Let $f\left(
y\right) =\sum_{j}\omega _{j}y^{\nu _{j}}$ where $\nu _{j}\in(0,1]$, $\omega
_{j}>0,j=1,...,n$ and $\beta >\frac{1}{2}\kappa ^{2}\left( 1-\nu _{j}\right)
,$ $j=1,...,n$. Then,%
\[
\sum_{j}\omega _{j}g\left( x\right) ^{\nu _{j}}=x,\ \ \ \ f^{\prime }\left(
y\right) =\sum_{j}\omega _{j}\nu _{j}y^{\nu _{j}-1},\ \ \ \ f^{\prime \prime
}\left( y\right) =\sum_{j}\omega _{j}\nu _{j}\left( \nu _{j}-1\right) y^{\nu
_{j}-2}
\]%
and%
\begin{align}\label{ex-6}\hs{3pc}
dX_{t} =&\left( \sum_{j}\omega _{j}\nu _{j}g\left( X_{t}\right) ^{\nu
_{j}-1}\left( \beta -\frac{1}{2}\kappa ^{2}\left( 1-\nu _{j}\right) \right)
-\alpha \sum_{j}\omega _{j}\nu _{j}g\left( X_{t}\right) ^{\nu _{j}}\right) dt
\\
&\;-\kappa \sum_{j}\omega _{j}\nu _{j}g\left( X_{t}\right) ^{\nu
_{j}-1/2}dB_{t}.\nonumber
\end{align}
The elasticity of variance
is a non-linear function of the VIX. The process has non-linear speed of
mean reversion and a non-linear attractor.
\end{example}

2. Here, we justify the relevance and choice of models based on the classes of functions $(A1)$ and $(A2)$ and the process $Y.$
As shown by Goard and Mazur (2013), the 3/2-model (Example 2.1) provides a better fit to the VIX data than various
alternatives including GBMP, MRGP, MRSRP and MRLP.
Notable features of this model are (i) a high power law of 3/2 which can reduce the heteroskedasticity of volatility and (ii) a nonlinear drift that generates substantial nonlinear mean-reverting behaviour when the volatility exceeds its long-run mean. Another important feature of this framework is that it reproduces the positive skew of implied volatilities which is the most relevant stylized fact of the VIX market, see, e.g., Mencia and Sentana (2013).
We note that all the models in Examples 2.2-2.3 exhibit this important property.

However, there are at least two reasons to consider generalizations of the 3/2 model.
Firstly, Goard and Mazur (2013) performed tests
for overidentifying restrictions relative to a specific benchmark which has a more general structure that nests the various alternatives
tested. Unconstrained GMM shows that the benchmark
has an estimated $\gamma $ of 1.48 and places large weights on the various
nonlinear components in the drift. These findings motivate us  to vary the power of the diffusion coefficient (Example 2.2) and
combine different powers (Example 2.3) in order to obtain a better fit to the VIX data.
Secondly, if one chooses parameters under a risk neutral measure to exactly match at-the-money vanilla options, then the 3/2 model generally undervalues
both in- and out-of-the money option prices, see, e.g., Section 7 in Goard and Mazur (2013). By taking models in Examples 2.2 and 2.3, one can adjust the tail behaviour of VIX either at 0 or at high levels and therefore improve model prices for in-the-money or out-of-the money vanillas compared to Example 2.1.

A thorough empirical analysis of the models introduced here is clearly needed. This is left for future research as the main aim of this paper is to provide a rigorous analysis of American options under these new specifications for VIX and numerical illustrations of the theoretical results. Nevertheless, based on the discussion above,  it is reasonable to introduce the generalized 3/2 and 1/2 models. Note also that building on Mencia and Sentana (2013), it might be useful to specify the parameter $\beta$ of $Y$ in \eqref{CIR} as a stochastic process instead of a constant in order to improve the fit to the VIX futures term structure. In this case, the American option pricing problem becomes a three-dimensional optimal stopping problem. This extension is left for future research as well.
\vs{6pt}

3. In this paper, we study the American VIX call and put options under the model \eqref{vix-1} with $f$ of types $(A1)$ and $(A2)$.
 By definition, the payoff of the American VIX call at exercise time $\tau\in[0,T]$ is $(X_\tau-K)^+ :=\max(X_\tau-K, 0)$ where $K>0$ is the strike and $T>0$ is the expiry date.
 The rational price $C^A$ of the American VIX call at time $t=0$ is the value function of the following optimal stopping problem
\begin{equation} \label{problem-1} \hs{6pc}
C^A=\sup \limits_{0\le\tau\le T}\EE e^{-r\tau} (X_\tau\m K)^+
 \end{equation}
where the supremum is taken over all stopping times $\tau$ of the process $X$, the expectation $\EE$ is taken under a risk neutral measure $\QQ$ and $r>0$ is the constant interest rate.

As the process $X$ is time-homogeneous Markov and \eqref{problem-1} is a finite horizon problem, we will study the problem \eqref{problem-1} in the Markovian setting and hence, we introduce dependence on time $t$ and the initial value of $X$
 \begin{equation} \label{problem-3} \hs{6pc}
C^A(t,x)=\sup \limits_{0\le\tau\le T-t}\EE e^{-r\tau}G(X^x_\tau)
 \end{equation}
 for $t\in [0,T)$ and $x>0$ where $X^x$ means that the process $X$ starts from $X^x_0=x$ and the payoff function $G$ is given by
 \begin{equation} \label{problem-4} \hs{9pc}
 G(x):=(x\m K)^+
 \end{equation}
for $x>0$. We tackle the problem \eqref{problem-3} in Sections 3 and 4. The discussion of the American put option follows in Section 5.
\vs{6pt}

4. Now we introduce the rational price function of the European VIX call option
\begin{equation} \label{problem-5} \hs{6pc}
C^E(t,x)=e^{-r(T-t)}\EE (X^x_{T-t}\m K)^+
 \end{equation}
 for $t\in [0,T)$ and $x>0$. A formula for \eqref{problem-5} in the $3/2$ model was derived by Goard and Mazur (2013)
 using the fact that the process $(1/X_t)_{t\ge 0}$  is a mean-reverting square-root process.
 We exploit a similar idea and recall that $X_t=f(Y_t)$ so that using the known probability density function
 $q(\wt{y};t,y)$ of $Y_t$, one can compute \eqref{problem-5} by numerical integration in an efficient way for $f$ of $3/2$ type as follows
\begin{equation} \label{Eur-1} \hs{6pc}
C^E(t,x)=e^{-r(T-t)}\int_0^{g(K)} (f(\wt{y})-K)\,q(\wt{y};T\m t,g(x))\,d\wt{y}
 \end{equation}
 for $t\in [0,T)$ and $x>0$ as $g$ is decreasing in this case. When $f$ is of $1/2$ type so that $g$ is increasing, the European price is
  \begin{align} \label{Eur-2} \hs{6pc}
C^E(t,x)=e^{-r(T-t)}\int_{g(K)}^\infty (f(\wt{y})-K)\,q(\wt{y};T\m t,g(x))\,d\wt{y}.
 \end{align}
\vs{6pt}

5. Below, we discuss how to compute efficiently the VIX futures term structure $F_{T}(x)=\EE[X^x_{T}]$  for any initial level $x>0$ of VIX and  different maturities $T>0$. Clearly, one can exploit efficient numerical integration
\begin{equation} \label{Fut-1} \hs{6pc}
F_{T}(x)=\int_0^{\infty} f(\wt{y})\,q(\wt{y};T,g(x))\,d\wt{y}
 \end{equation}
for $f$ of either $3/2$ or $1/2$ type. Moreover, it is also possible to obtain a closed-form approximation of the futures price
\begin{equation} \label{Fut-2} \hs{2pc}
F_{T}=\EE[X_T]=\EE[f(Y_T)]=f(y)+\sum_{k=1}^\infty \frac{1}{k!}\EE (Y_T-y)^k f^{(k)}(y)
 \end{equation}
 and if we choose $y=\EE [Y_T]$ with 4-th order of approximation then
 \begin{equation} \label{Fut-3} \hs{2pc}
F_{T}\approx f(\EE[Y_T])+\sum_{k=2}^4  \frac{1}{k!}\EE(Y_T-\EE[Y_T])^k f^{(k)}(\EE[Y_T])
 \end{equation}
 where the centered moments of $Y_T$ are well known as it has a non-central chi-squared distribution. This approximation shows good performance and the error is usually bounded by 1\%. It can be further improved by taking higher order in the Taylor series. However, if one needs accurate values, then the numerical integration of \eqref{Fut-1} should be used and it is quite fast.
 \vs{6pt}

 \section{The free-boundary problem for the American VIX call option}

 In this section we will reduce the problem \eqref{problem-3} to a free-boundary problem and the latter will be tackled in the next section using the local time-space calculus (see Peskir (2005a)).
 First, using that the payoff function $G(x)$ is continuous and standard arguments (see e.g. Corollary 2.9 (Finite horizon) with Remark 2.10 in Peskir and Shiryaev (2006)), we have that the continuation and exercise regions read, respectively
\begin{align} \label{C} \hs{5pc}
&\cC= \{\, (t,x)\in[0,T)\! \times\! [0,\infty):C^A(t,x)>G(x)\, \} \\[3pt]
 \label{D}&\cE= \{\, (t,x)\in[0,T)\! \times\! [0,\infty):C^A(t,x)=G(x)\, \}
 \end{align}
and the optimal stopping time in \eqref{problem-3} is given by
\begin{align} \label{OST} \hs{5pc}
\tau=\inf\ \{\ 0\leq s\leq T-t:(t\p s,X^x_{s})\in\cE\ \}.
 \end{align}

 Before starting our analysis, we recall an important result for our purposes on flows of stochastic differential equations.
The underlying model satisfies the conditions of Theorem 37 of Chapter V, Section 7  in Protter (1990), i.e., which simply requires only locally Lipschitz coefficients for the SDE \eqref{vix-1},
so that we have the following inequality
\begin{equation} \label{flow} \hs{6pc}
\left[\EE \sup\limits_{0\leq u\leq T-t}\left(X^x_u\m X^y_u\right)^2\right]^{1/2}\le C_L \left|x-y\right|
 \end{equation}
for $x,y>0$ and some constant $C_L>0$. We will use this estimate for the proof of the smooth-fit property.
\vs{6pt}

1. We show that the price function $C^A$ is continuous on $[0,T)\times (0,\infty)$.
It follows that
\begin{align} \label{cont-1} \hs{1pc}
 0\le&\; C^A(t,x)-C^A(t,y)\le\sup \limits_{0\leq\tau\leq T-t}\EE e^{-r\tau}\left(X^x_\tau\m X^y_\tau\right)
  \le \EE \sup\limits_{0\leq u\leq T-t}\left(X^x_u\m X^y_u\right)\\
 \le& \left(\EE \sup\limits_{0\leq u\leq T-t}\left(X^x_u\m X^y_u\right)^2\right)^{1/2}
 \le C_L (x\m y)\nonumber
 \end{align}
for $x\ge y$ and $t\in [0,T)$ where we used that $\sup(f)-\sup(g)\leq\sup(f\m g)$ and
$(x-K)^{+}-(y-K)^{+}\leq(x-y)^{+}$ for $x,y,K\in \R$, the  comparison theorem for solutions
of SDEs (i.e. $\QQ(X^x_s\ge X^y_s, \; s\ge 0)=1$), Holder inequality and the inequality \eqref{flow} . From \eqref{cont-1} we see that $x\mapsto C^A(t,x)$ is continuous uniformly over $t\in [0,T]$. Thus to prove that $C^A$ is continuous on $[0,T)\times(0,\infty)$, it is enough to show that $t\mapsto C^A(t,x)$ is continuous on $[0,T]$ for each $x>0$ given and fixed. For this, take any $t_1<t_2$ in $[0,T]$
and let $\tau_1$ be an optimal stopping time for $C^A(t_1,x)$.  Setting $\tau_2=\tau_1\wedge (T-t_2)$ and using that
$t\mapsto C^A(t,x)$ is decreasing on $[0,T]$, we have
\begin{align} \label{cont-2} \hs{1pc}
0\le C^A(t_1,x)-C^A(t_2,x)\le \EE e^{-r\tau_1} G(X^x_{\tau_1})-\EE e^{-r\tau_2} G(X^x_{\tau_2})
\le \EE \left(X^x_{\tau_1}\m X^x_{\tau_2}\right)^+.
\end{align}
Letting first $t_2-t_1\rightarrow0$ and using $\tau_1-\tau_2\rightarrow0$, we see that $C^A(t_1,x)-C^A(t_2,x)\rightarrow0$ by dominated convergence. This shows that
$t\mapsto C^A(t,x)$ is continuous on $[0,T]$, and the proof of the initial claim is complete.
\vs{6pt}

2. Now we get some initial insights into the structure of exercise region $\cE$.
\vs{2pt}

$(i)$ We first calculate the function $H(x)\!:=(\L_X G \m rG)(x)$ for $x\in(0,\infty)$ (which is the instantaneous benefit
of waiting to exercise) where
\begin{align} \label{H-0} \hs{3pc}
\L_X =\Big(f'(g(x))\big(\beta\m\alpha g(x)\big)+\frac{1}{2}\kappa^2 f''(g(x))g(x)\Big)\frac{d}{dx}+\frac{1}{2}\kappa^2 (f'(g(x)))^2 g(x)\,  \frac{d^2}{dx^2}
\end{align}
is the infinitesimal generator of $X$.
As $G(x)=(x\m K)^+$, we have  that
\begin{align} \label{H-1} \hs{6pc}
H(x)=h(x)I(x\ge K)
\end{align}
for $x\in(0,\infty)$ where
\begin{align} \label{H-2} \hs{3pc}
h(x)=f'(g(x))\big(\beta\m\alpha g(x)\big)+\frac{1}{2}\kappa^2 f''(g(x))g(x)-r(x\m K)
\end{align}
for $x>0$.
Throughout the paper, the following condition is imposed on the model and we note that all models in Examples 2.1-2.6 satisfy this assumption (the verification is provided in the Appendix):
\vs{2pt}

\noindent \textbf{Assumption R}: There exists $x^{\ast}>0$ such that
$h(x) \ge 0$ if and only if $x\le x^{\ast}$.
\vs{2pt}

We could assume a weaker condition, that there exists $x^*>0$ such that $H(x)\ge 0$ if and only if $x\le \max(K,x^*)$. We use \textbf{Assumption R} in order to have a unified condition
for both call and put options, and it is enough for models of interest such as Examples 2.1-2.6.
\vs{6pt}

$(ii)$  We now use the Ito-Tanaka's formula and the definition of $H$ to obtain
\begin{align} \label{Tanaka-1} \hs{3pc}
 \EE e^{-r\tau}G(X^{x}_\tau)=\;G(x)+\EE \int_0^\tau e^{-rs}H(X^{x}_s)ds+\frac{1}{2}\EE \int_0^\tau e^{-rs}d\ell^{K}_s (X^x)
 \end{align}
for $x\in(0,\infty)$ and any stopping time $\tau$ of the process $X$ where $(\ell^K_s(X))_{s\ge 0}$ is the local time process of $X$ at level $K$
\begin{align} \label{Tanaka-2} \hs{3pc}
 \ell^{K}_s (X^x):=\QQ-\lim_{\eps \downarrow 0}\frac{1}{2\eps}\int_0^{s} I(K\m\eps<X^x_u<K\p\eps)d\left \langle X,X \right \rangle_u
 \end{align}
 and $d\ell^{K}_s (X^x)$ refers to the integration with respect to the continuous increasing function $s\mapsto \ell^{K}_s (X^x)$.
The equation \eqref{Tanaka-1} and \textbf{Assumption R} show that it is not optimal to exercise the call option when $X_t\le \max(K,x^*)$ as $H(X_t)\ge0$ in this region and thus both integral terms
on the right-hand side of \eqref{Tanaka-1} are non-negative. This fact can be also explained in the particular case where $X_t<K$ as follows:
by exercising below $K$, the option holder receives a null payoff, whereas  by waiting would have a positive probability of collecting a strictly positive payoff in the future.
\vs{2pt}

Another implication of \eqref{Tanaka-1} is that the exercise region is non-empty for all $t\in[0,T)$, as for large $x\uparrow \infty$ the integrand $H$
is negative and the local time term is zero, and thus due to a lack of time to compensate for the negative $H$, it is optimal to stop at once.
\vs{6pt}

3. Next we prove further properties of the exercise region $\cE$ and define the optimal exercise boundary.

$(i)$ As the payoff function in \eqref{problem-3} is time-independent,
it follows that the map $t\mapsto C^A (t,x)$ is non-increasing on $[0,T]$ for each $x>0$ so that $C^A(t_1,x)\m G(x)\ge C^A(t_2,x)\m G(x)\ge0$
for $0\le t_1 <t_2 <T $ and $x\in(0,\infty)$. Now, if we take a point $(t_1,x)\in \cE$, i.e. $C^A(t_1,x)=G(x)$, then $(t_2,x)\in \cE$ as well, which shows that the exercise region is increasing in $t$. In other words, $\cE$ is right-connected.
\vs{2pt}

$(ii)$ Now let us take $t>0$ and $x>y>\max(K,x^*)$ such that $(t,y)\in \cE$.
Then, by right-connectedness of the exercise region, we have that $(s,y)\in \cE$ as well for any $s>t$. If we now run the process $(s,X_{s-t})_{s\ge t}$ from $(t,x)$, we cannot hit the level $\max(K,x^*)$ before exercise (as $x>y$), thus the local time term in \eqref{Tanaka-1} is 0 and integrand $H$ is negative (by \textbf{Assumption R}).
Therefore, it is optimal to exercise at $(t,x)$ and we get up-connectedness of the exercise region $\cE$.
\vs{2pt}

$(iii)$ From $(i)$-$(ii)$ and paragraph 2$(ii)$ above, we can conclude that
there exists an optimal exercise boundary $b:[0,T]\rightarrow (0,\infty)$ such that
\begin{align} \label{OST-2} \hs{5pc}
&\tau_b=\inf\ \{\ 0\leq s\leq T\m t:X^x_{s}\ge b(t\p s) \ \}
 \end{align}
is optimal in \eqref{problem-3} and $\max(K,x^*)<b(t)<\infty$ for $t\in[0,T)$. Moreover, $b$ is decreasing on $[0,T)$.

\begin{remark}
If \textbf{Assumption R} does not hold and the function $h(x)$ changes sign more than once for $x>K$, then there are more than one exercise boundary.
Therefore the exercise region $\cE$ is disconnected.
\end{remark}

4. Now we prove that the smooth-fit condition along the boundary $b$ holds
 \begin{align}\label{SF}\hs{4pc}
C^A_x (t, b(t)-)=C^A_x(t,b(t)+)=G'(b(t))=1
\end{align}
for all $t\in[0,T)$. To the best of our knowledge, in the literature on optimal stopping problems, the smooth-fit property has been proven in models where the  dependence of $X^x$ on $x$ is given  explicitly
(e.g. Brownian motion or geometric Brownian motion),
however in our model such dependence is unknown. For this reason, we provide another proof based on
the inequality \eqref{flow}.

$(i)$ First let us fix a point $(t,x)\in [0,T)\times(0,\infty)$ lying on the boundary $b$ so that $x=b(t)$.
Then, we have
\begin{align} \label{SF-1} \hs{4pc}
\frac{C^A(t,x)-C^A(t,x\m\eps)}{\eps}&\le \frac{G(x)-G(x\m\eps)}{\eps}
\end{align}
and taking the limit as $\varepsilon\downarrow 0$, we get
\begin{align} \label{SF-2} \hs{6pc}
C^A_x (t,x-)\le G'(x)=1
\end{align}
 where the left-hand derivative exists by monotonicity of $x\mapsto C^A(t,x)$ on $(0,\infty)$ for any fixed $t\in[0,T)$.
\vs{2pt}

$(ii)$ To prove the reverse inequality, we set $\tau_\eps=\tau_\eps(t,x\m\eps)$ as an optimal stopping time for $C^A(t,x-\eps)$.
Using that $X$ is a regular diffusion  and $t\mapsto b(t)$ is decreasing,  we have that
$\tau_\eps\to0$ as $\eps\to0$ $\QQ$-a.s. By the comparison theorem for solutions
of SDEs and noting that %
\begin{align}\label{SF-3a}\hs{+2pc}
G( X_{\tau _{\eps }}^{x}&) \m G( X_{\tau _{\eps}}^{x-\eps })  \\
=&\left( X_{\tau _{\eps }}^{x}\m X_{\tau
_{\eps }}^{x-\eps }\right) I(X_{\tau _{\eps }}^{x-\eps
}\geq K)+(X_{\tau _{\eps }}^{x}\m K)I(X_{\tau _{\eps }}^{x}\geq K\geq
X_{\tau _{\eps }}^{x-\eps }) \nonumber\\
\geq &\left( X_{\tau _{\eps }}^{x}\m X_{\tau _{\eps }}^{x-\eps
}\right) I(X_{\tau _{\eps }}^{x-\eps }\geq K)\nonumber
\end{align}%
we obtain
 \begin{align}\label{SF-3}\hs{+2pc}
\frac{1}{\eps}\Big(&C^A(t,x)-C^A(t,x\m\eps)\Big)\\
&\ge \frac{1}{\eps}
\EE \left[e^{-r\tau_{\eps}}\left( X_{\tau _{\eps }}^{x}\m X_{\tau _{\eps }}^{x-\eps
}\right) I(X_{\tau _{\eps }}^{x-\eps }\geq K)\right]\nonumber\\
&=\frac{1}{\eps}\EE \left[e^{-r\tau_{\eps}}\left( X_{\tau _{\eps }}^{x}\m X_{\tau _{\eps }}^{x-\eps
}\right)\right]-\frac{1}{\eps}\EE \left[e^{-r\tau_{\eps}}\left( X_{\tau _{\eps }}^{x}\m X_{\tau _{\eps }}^{x-\eps
}\right) I(X_{\tau _{\eps }}^{x-\eps }\leq K)\right]\nonumber.
\end{align}
 Then the second term on the right-hand side of
\eqref{SF-3} goes to 0 as $\eps\to0$ as
 \begin{align}\label{SF-4}\hs{2pc}
0&\le \frac{1}{\eps}\EE \left[e^{-r\tau_\eps}\left( X_{\tau _{\eps }}^{x}\m X_{\tau _{\eps }}^{x-\eps
}\right) I(X_{\tau _{\eps }}^{x-\eps }\leq K)\right]\\
&\le \frac{1}{\eps}\left(\EE \left( X_{\tau _{\eps }}^{x}\m X_{\tau _{\eps }}^{x-\eps
}\right)^2\right)^{1/2} \left(\QQ(X_{\tau _{\eps }}^{x-\eps }\leq K)\right)^{1/2}\nonumber\\
&\le \frac{1}{\eps}\left(\EE \sup\limits_{0\leq u\leq T-t}\left( X_{u}^{x}\m X_{u}^{x-\eps
}\right)^2\right)^{1/2} \left(\QQ(X_{\tau _{\eps }}^{x-\eps }\leq K)\right)^{1/2}\nonumber\\
&\le C_L \left(\QQ(X_{\tau _{\eps }}^{x-\eps }\leq K)\right)^{1/2}\rightarrow 0\nonumber
\end{align}
where we used the Holder inequality, the inequality \eqref{flow} and that the latter probability goes to zero because $x>K$.
 Now we turn to the first term on the right-hand side of \eqref{SF-3}. Using Ito's formula, we have:
 \begin{align}\label{SF-5}\hs{2pc}
\frac{1}{\eps}\EE \left[e^{-r\tau_\eps}\left( X_{\tau _{\eps }}^{x}\m X_{\tau _{\eps }}^{x-\eps
}\right)\right]=1+\frac{1}{\eps}
\EE \left[\int_0^{\tau_\eps} e^{-rs}\left(\omega(X^x_s) \m \omega(X^{x-\eps}_s)\right)ds\right]
\end{align}
where $\omega(x):=f'(g(x))\big(\beta\m\alpha g(x)\big)\p\frac{1}{2}\kappa^2 f''(g(x))g(x)\m rx$ for $x>0$. We show that the second term of \eqref{SF-5} goes to 0 as $\eps\to0$
\begin{align}\label{SF-6}
0&\le \frac{1}{\eps}
\EE \left|\int_0^{\tau_\eps} e^{-rs}(\omega(X^x_s) \m \omega(X^{x-\eps}_s))ds
\right|\le \frac{1}{\eps}
\left[\EE \int_0^{\tau_\eps} e^{-rs}|\omega'(\xi_s)|(X^x_s \m X^{x-\eps}_s)ds\right]\\
&\le \frac{1}{\eps}
\EE\left[ \int_0^{\tau_\eps} |\omega'(\xi_s)|ds\cdot \sup\limits_{0\leq u\leq T-t}\left( X_{u}^{x}\m X_{u}^{x-\eps
}\right)\right]\nonumber\\
&\le \frac{1}{\eps}  \left(\EE\left[ \int_0^{\tau_\eps} |\omega'(\xi_s)|ds\right]^2\right)^{1/2}\left(\EE \sup\limits_{0\leq u\leq T-t}\left( X_{u}^{x}\m X_{u}^{x-\eps
}\right)^2 \right)^{1/2}\nonumber\\
&\le \frac{1}{\eps} C_L\, \eps  \left(\EE\left[ \int_0^{\tau_\eps} |\omega'(\xi_s)|ds\right]^2\right)^{1/2}= C_L  \left(\EE\left[ \int_0^{\tau_\eps} |\omega'(\xi_s)|ds\right]^2\right)^{1/2} \nonumber
\end{align}
where we used the mean value theorem and choice $\xi_s\in[X^{x-\eps}_s,X^{x}_s]$, then H\"older inequality and inequality \eqref{flow}.
Now we show that $\int_0^{\tau_\eps} |\omega'(\xi_s)|ds\rightarrow 0$ as $\eps\downarrow 0$ $\QQ$-a.s. Indeed, let us fix the sample path of $B$ and take an arbitrary $\delta>0$. Then, for some $\eps_0<x$ and for the corresponding trajectory of $X^{x-\eps_0}_s$ we define
$t'>0$ as the first exit time from the compact set $[x',x]$ for fixed $x'<x-\eps_0$.
Thus, the values of $X^{x-\eps_0}_s$ belong to this compact set  for $s\in[0,t']$. Next, by the locally Lipschitz continuity of $\omega'$, we can bound $|\omega'(\cdot)|$ from above on this compact set by some constant $C_{\omega'}$.
As $\tau_\eps\rightarrow 0$ $\QQ$-a.s, we then choose $\eps'<\eps_0$ small enough such that for $\eps<\eps'$ we have $\tau_\eps<\min(t',\delta/C_{\omega'})$. By the comparison theorem for SDEs and as $b$ is decreasing,
$X^{x-\eps}$ belongs to the same compact set before $\tau_\eps<t'$. Therefore  we
have that $\int_0^{\tau_\eps} |\omega'(\xi_s)|ds\le C_{\omega'} \tau_\eps<\delta$ for $\eps<\eps'$ and it follows that $\int_0^{\tau_\eps} |\omega'(\xi_s)|ds\rightarrow 0$ as $\eps\downarrow 0$ $\QQ$-a.s.
Then,
we obtain that $\EE\left[ \int_0^{\tau_\eps} |\omega'(\xi_s)|ds\right]^2$ as
$\eps\to0$ by the monotone convergence theorem.

 Thus, using \eqref{SF-3}-\eqref{SF-6} and taking the limits as $\eps\to0$, we have that
\begin{align} \label{SF-7} \hs{5pc}
C^A_x (t,x-)\ge G'(x)=1
\end{align}
for $t\in[0,T)$. Thus, combining \eqref{SF-2} and \eqref{SF-7}, we obtain \eqref{SF}.
\vs{6pt}

5. Here, we prove that the boundary $b$ is continuous on $[0,T]$ and that $b(T-)=\max(K,x^*)$.
The proof is provided in 3 steps and follows the approach proposed by De Angelis (2014).
\vs{+2pt}

$(i)$ We first show that $b$ is right-continuous. Let us fix $t\in [0,T)$ and take a
sequence $t_n \downarrow t$ as $n\rightarrow \infty$. As $b$ is decreasing, the right-limit $b(t+)$ exists and
$(t_n, b(t_n))$ belongs to $\cE$ for all $n\ge 1$. Recall that $\cE$ is closed so that $(t_n,b(t_n))\to (t,b(t+))\in \cE$ as $n\to \infty$ and we may conclude that $b(t+)\ge b(t)$. The fact that $b$
is decreasing gives the reverse inequality and thus $b$ is right-continuous as claimed.
\vs{+2pt}

$(ii)$ Now we prove that $b$ is also left-continuous. Assume that there exists $t_0\in(0,T)$ such that $b(t_0-)>b(t_0)$.
Let us set $x_1 = b(t_0)$ and $x_2 = b(t_0-)$ so that $x_1 < x_2$. For $\eps \in (0,(x_2\m x_1)/2)$ given and fixed, let $\varphi_\eps : (0,\infty) \rightarrow [0,1]$ be a $C^\infty$- function satisfying (i) $\varphi_\eps (x) = 1$ for $x\in [x_1 \p\eps, x_2 \m\eps]$ and
(ii) $\varphi_\eps (x) = 0$ for $x\in (0,x_1\p \eps/2]\cup [x_2\m\eps/2,\infty)$. Letting $\L^*_X $ denote the adjoint of
$\L_X$, recalling that $t\rightarrow  C^A(t,x)$ is decreasing on $[0,T]$ and that $C^A_t\p \L_X C^A  \m rC^A= 0$ on $\cC$, we find integrating by parts (twice) that
\begin{align}\label{cont01}\hs{1pc}
0\ge \int^{x_2}_{x_1}{\varphi_\eps(x)C^A_t(t_0 \m \delta,x)dx}=-\int^{x_2}_{x_1}{C^A(t_0 \m \delta,x)\left(\L_X^*\varphi_\eps(x)\m r\varphi_\eps(x)\right)dx}
\end{align}
for $\delta\in (0,t_0\wedge  (\eps/2))$ so that $\varphi_\eps (x_2\m\delta) = \varphi'_\eps (x_2\m\delta)  = 0$ as needed. Letting $\delta \downarrow 0$, it follows using the dominated convergence theorem and integrating by parts (twice) that
\begin{align}\label{cont02}\hs{1pc}
0&\ge -\int^{x_2}_{x_1}{C^A(t_0,x)\left(\L_X^*\varphi_\eps(x)\m r\varphi_\eps(x)\right)dx}
=-\int^{x_2}_{x_1}{G(x)\left(\L_X^*\varphi_\eps(x)\m r\varphi_\eps(x)\right)dx}\\
&=-\int^{x_2}_{x_1}{\left(\L_X G(x)\m rG(x)\right)\varphi_\eps(x) dx}=-\int^{x_2}_{x_1}{H(x)\varphi_\eps(x)dx}.\nonumber
\end{align}
Letting $\eps\downarrow  0$, we obtain
\begin{align}\label{cont03}\hs{5pc}
0\ge -\int^{x_2}_{x_1}{H(x)dx}>0
\end{align}
as $x\rightarrow H(x)$ is strictly negative on $(x_1,x_2]$. We thus have a contradiction and therefore we may conclude that $b$ is continuous on $[0, T )$ as claimed.
\vs{+2pt}

$(iii)$ To prove that $b(T-)=\max(K,x^*)$, we can use the same arguments as those in $(ii)$ above with $t_0=T$ and suppose that $b(T-)>\max(K,x^*)$.

\vs{6pt}

6. The facts proved in paragraphs 1-5 above and  standard arguments based on the strong Markov property (see, e.g., Peskir and Shiryaev (2006)) lead to the following free-boundary problem for the value function $C^A$ and unknown boundary $b$
\begin{align} \label{PDE} \hs{5pc}
&C^A_t \p\L_X C^A \m rC^A=0 &\hs{-30pt}\text{in}\;  \cC\\
\label{IS}&C^A(t,b(t))=G(b(t))=b(t)\m K &\hs{-30pt}\text{for}\; t\in[0,T)\\
\label{SP}&C^A_x (t,b(t))=G'(b(t))=1 &\hs{-30pt}\text{for}\; t\in[0,T) \\
\label{FBP1}&C^A(t,x)>G(x) &\hs{-30pt}\text{in}\; \cC\\
\label{FBP2}&C^A(t,x)=G(x) &\hs{-30pt}\text{in}\; \cE
\end{align}
where the continuation set $\cC$ and the exercise set $\cE$ are given by
\begin{align} \label{C-1} \hs{5pc}
&\cC= \{\, (t,x)\in[0,T)\! \times\! (0,\infty):x<b(t)\, \} \\[3pt]
 \label{D-1}&\cE= \{\, (t,x)\in[0,T)\! \times\! (0,\infty):x\ge b(t)\, \}.
 \end{align}
The following properties of $C^A$ and $b$ were also verified above
\begin{align} \label{Prop-1} \hs{5pc}
&C^A\;\text{is continuous on}\; [0,T]\times(0,\infty)\\
\label{Prop-2}&C^A\;\text{is}\; C^{1,2}\;\text{on}\; \cC\\
\label{Prop-3}&x\mapsto C^A(t,x)\;\text{is increasing on $[0,\infty)$ for each $t\in[0,T]$}\\
\label{Prop-4}&t\mapsto C^A(t,x)\;\text{is decreasing on $[0,T]$ for each $x\in [0,\infty)$}\\
\label{Prop-5}&t\mapsto b(t)\;\text{is decreasing and continuous on $[0,T]$ with}\; b(T-)=\max(K,x^*).
\end{align}

\vs{6pt}

\section{The rational price of the American VIX call option}

 We will show in this section that the optimal exercise boundary $b$ can be obtained as the unique solution to a nonlinear integral equation of Volterra type. We then provide the early exercise premium representation formula for the rational price $C^A$, which decomposes it into the sum of the  European VIX call price $C^E$
 and the early exercise premium which depends on the exercise boundary $b$.
\vs{+6pt}

1. We recall that we already showed how to compute the European call price in Section 2 above. Now we denote the following function
\begin{align} \label{L} \hs{5pc}
L(u,x,z)=-\EE \big[e^{-ru}H(X^x_u) I(X^x_u \ge z)\big]
 \end{align}
 for $u\ge 0$ and $x,z>0$.
 This function can be computed using the same idea as for the European call price.
  Using that
$Y_t=g(X^x_t)$ is the mean-reverting square-root process and that the random variable $Y^y_t$ has non-central chi-squared density function
 $q(\wt{y};t,y)$, we have for $f$ of $3/2$- type that
 \begin{align} \label{L-3} \hs{5pc}
L(u,x,z)&=-\EE \Big[e^{-ru}H\big(f(Y^{g(x)}_u)\big) I(Y^{g(x)}_u \le g(z))\Big]\\
&=-e^{-ru}\int_0^{g(z)} H(f(\wt{y}))\,q(\wt{y};u,g(x))\,d\wt{y}\nonumber
 \end{align}
 for $u\ge 0$ and $x,z>0$ as $g$ is decreasing in this case. When $f$ is of $1/2$- type so that $g$ is increasing, the function $L$ can be computed as
  \begin{align} \label{L-4} \hs{5pc}
L(u,x,z)&=-\EE \Big[e^{-ru}H\big(f(Y^{g(x)}_u)\big) I(Y^{g(x)}_u \ge g(z))\Big]\\
&=-e^{-ru}\int_{g(z)}^\infty H(f(\wt{y}))\,q(\wt{y};u,g(x))\,d\wt{y}\nonumber
 \end{align}
for $u\ge 0$ and $x,z>0$.
  \vs{+6pt}

 2. The main result of this section can now be stated as follows.
\begin{theorem}\label{th:1}
The price function $C^A$ in \eqref{problem-3} has the representation
\begin{align}\label{th-1} \hs{4pc}
C^A(t,x)=\;C^E(t,x) +\int_0^{T-t}L(u,x,b(t\p u))du
\end{align}
for $t\in[0,T]$ and $x\in (0,\infty)$. The optimal exercise boundary $b$ in \eqref{problem-3} can be characterized as the unique solution to the nonlinear integral equation of Volterra type
\begin{align}\label{th-2} \hs{4pc}
b(t)-K=\;C^E(t,b(t)) +\int_0^{T-t} L(u,b(t),b(t\p u))du
\end{align}
for $t\in[0,T]$, in the class of continuous decreasing functions $t\mapsto b(t)$ with $b(T)=\max(K,x^*)$ (See Figures 1 and 2).
\end{theorem}

\begin{figure}[t]
\begin{center}
\includegraphics[scale=0.75]{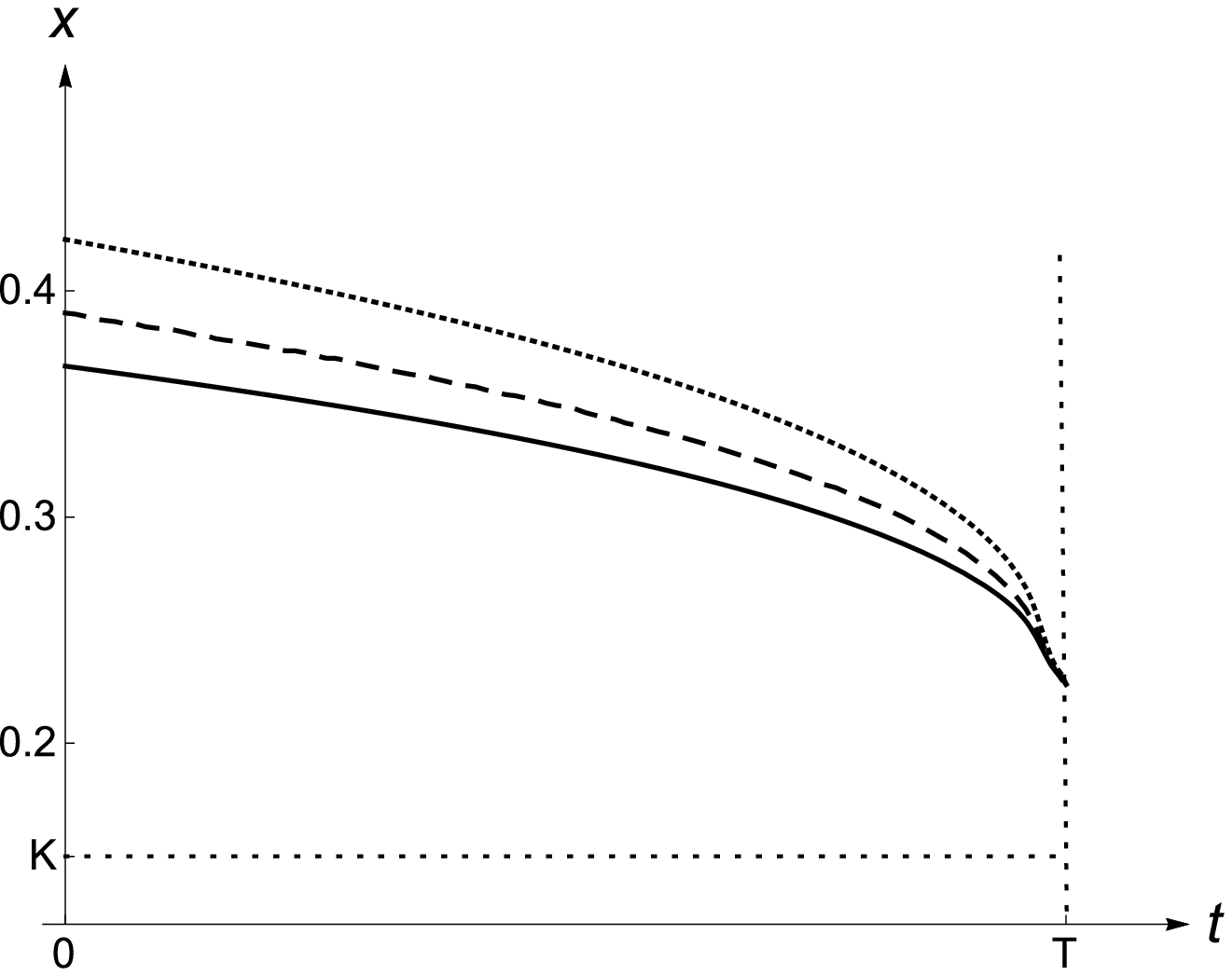}
\end{center}

{\par \leftskip=1.6cm \rightskip=1.6cm \small \ni \vs{-10pt}

\textbf{Figure 1.} This figure plots the optimal exercise boundary $b$ for models of $3/2$-type: $3/2$-model (solid), $1\p1/(2\nu)$-model with $\nu=1.2$ (dotted),
  and mixture $(3/2, 1\p 1/(2\nu))$-model with $\nu=1.2$ and weights $w_j=0.5$ (dashed). The parameter set is $T=1, K=0.15, r=0.05$, and the coefficients for the $3/2$-model are $\alpha=2.94, \beta=17.10, \kappa=2.05$.
For the $1\p1/(2\nu)$-model, the coefficient $\alpha=3.64$ is adjusted to have the same attractor as for the $3/2$-model;
for the mixture model $\alpha=3.27$. The parameters for the $3/2$-model were calibrated by Goard and Mazur (2013).

\par} \vs{10pt}

\end{figure}

\begin{proof}

$(A)$ First, we clearly have that the following conditions hold:
$(i)$ $C^A$ is $C^{1,2}$ on $\cC\cup \cE$;
$(ii)$ $b$ is of bounded variation (due to monotonicity);
 $(iii)$ $C^A_t\p\L_{X}C^A -rC^A$ is locally bounded;
 $(iv)$ $C^A_{xx}=F_1+F_2$ on $\cC\cup \cE$, where $F_1$ is non-negative and $F_2$ is continuous on $[0,T)\times(0,\infty)$;
$(v)$ $t\mapsto C^A_x (t,b(t)\pm)$ is continuous (recall \eqref{SP}). Hence, we can apply the local time-space formula on curves (Peskir (2005a)) for $e^{-rs}C^A(t\p s,X^x_s)$
\begin{align} \label{th-3} \hs{1pc}
e^{-rs}C^A&(t\p s,X^x_s)\\
=\;&C^A(t,x)+M_s\nonumber\\
 &+ \int_0^{s}
e^{-ru}\left(C^A_t \p\L_X C^A
\m rC^A\right)(t\p u,X^x_u)
 I(X^x_u \neq b(t\p u))du\nonumber\\
 &+\frac{1}{2}\int_0^{s}
e^{-ru}\left(C^A_x (t\p u,X^x_u +)-C^A_x (t\p u,X^x_u -)\right)I\big(X^x_u=b(t\p u)\big)d\ell^{b}_u(X^x)\nonumber\\
  =\;&C^A(t,x)+M_s + \int_0^{s}
e^{-ru}\left(\L_X G-rG\right)(t\p u,X^x_u)I(X^x_u \ge b(t\p u))du\nonumber\\
  =\;&C^A(t,x)+M_s +\int_0^{s}
e^{-ru}H(X^x_u) I(X^x_u \ge b(t\p u))du\nonumber
  \end{align}
where we used \eqref{PDE}, the smooth-fit condition \eqref{SP}, \eqref{FBP2} and where $M=(M_s)_{s\ge 0}$ is the martingale part and
 $(\ell^{b}_t(X^x))_{t\ge 0}$ is the local time process of $X^x$ at the boundary $b$
\begin{align} \label{Tanaka-3} \hs{3pc}
 \ell^{b}_t (X^x):=\QQ-\lim_{\eps \downarrow 0}\frac{1}{2\eps}\int_0^{t} I(b(t\p u)\m\eps<X^x_u<b(t\p u)\p\eps)d\left \langle X,X \right \rangle_u.
 \end{align}
 Now
upon letting $s=T-t$, taking the expectation $\EE$, recalling the definition of $C^E$ in \eqref{problem-5}, using the optional sampling theorem for $M$, rearranging terms and noting that
$C^A(T,x)=G(x)=(x\m K)^+$ for all $x>0$, we get \eqref{th-1}.
The integral equation \eqref{th-2} is obtained by inserting $x=b(t)$ into \eqref{th-1} and using \eqref{IS}.
\vs{6pt}

$(B)$ Now we show that $b$ is the unique solution to the equation \eqref{th-2} in the class of continuous functions $t\mapsto b(t)$.  We note that monotonicity and the terminal value $b(T)$ are not needed for uniqueness, we require only that $b\ge \max(K,x^*)$ on $[0,T)$. The proof is divided in several steps and is based on arguments similar to those employed by Du Toit and Peskir (2007) and originally derived by Peskir (2005b).

\begin{figure}[t]
\begin{center}
\includegraphics[scale=0.6]{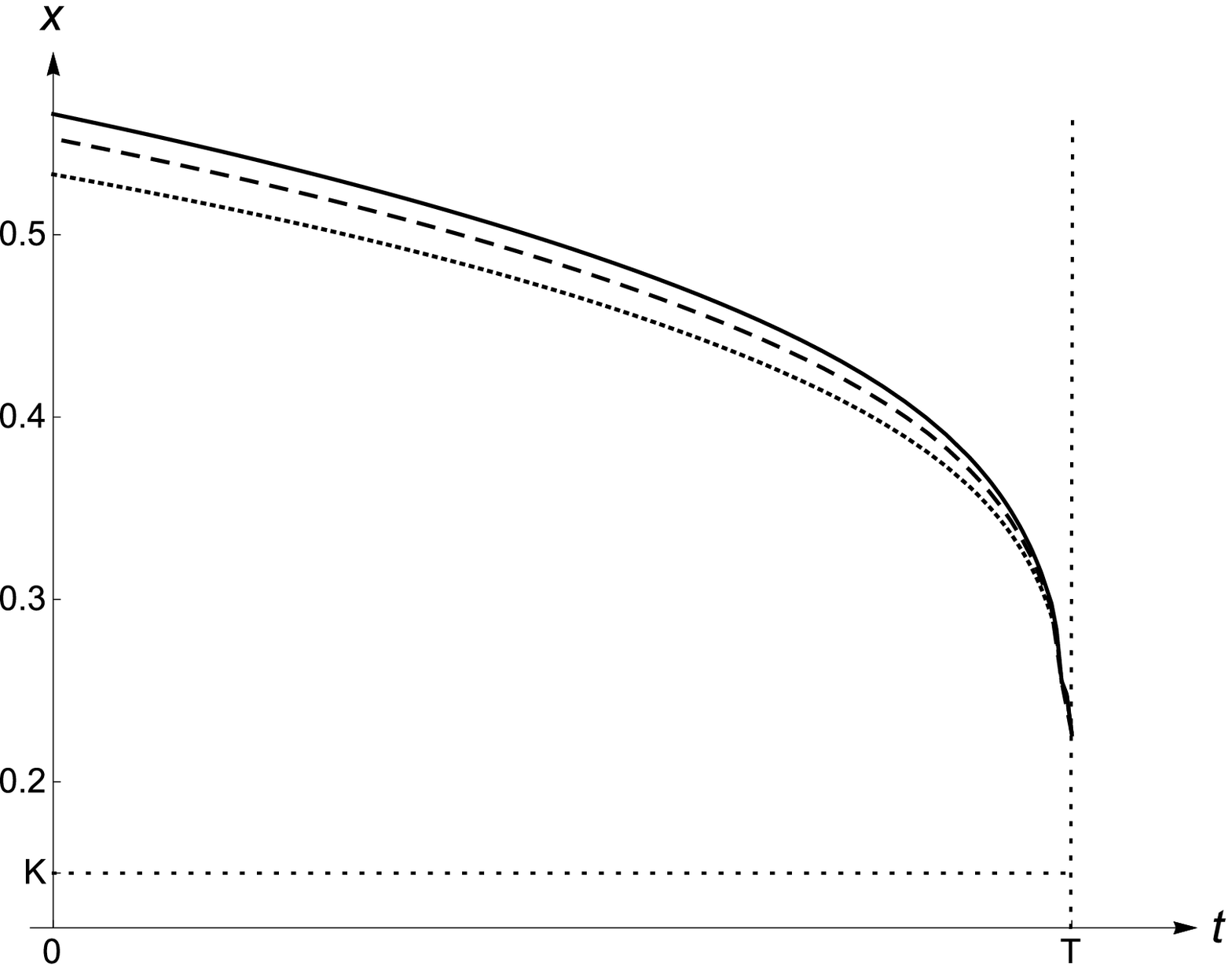}
\end{center}

{\par \leftskip=1.6cm \rightskip=1.6cm \small \ni \vs{-10pt}

\textbf{Figure 2.} This figure plots the optimal exercise boundary $b$ for models of $1/2$-type: $1/2$-model (solid), $1\m1/(2\nu)$-model with $\nu=0.8$ (dotted),
  and mixture $(1/2, 1\m1/(2\nu))$-model with $\nu=0.8$ and weights $w_j=0.5$ (dashed). The parameter set is $T=1$, $K=0.15$, $r=0.05$, and the coefficients for the $1/2$-model are $\alpha=3$, $\beta=0.68$, $\kappa=1$.
For the $1\m1/(2\nu)$-model the coefficient $\alpha=3.7$ is adjusted to have the same attractor as for the $1/2$-model;
for the mixture model $\alpha=2.9$.

\par} \vs{10pt}

\end{figure}

$(B.1)$ Let $c:[0,T]\rightarrow \R$  be a solution to equation
\eqref{th-2} such that $c$ is continuous decreasing with $c(T)=\max(K,x^*)$. We will show that $c$ must be equal to the optimal exercise boundary $b$.
Now let us consider the function $U^{c}:[0,T)\rightarrow \R$ defined as follows
\begin{align}\label{th-4} \hs{3pc}
U^{c}(t,x)=\;C^E(t,x)+\int_0^{T-t} L(u,x,c(t\p u))du
\end{align}
for $(t,x)\in[0,T]\times(0,\infty)$. Observe the fact that $c$ solves the equation
\eqref{th-2} means exactly that $U^{c}(t,c(t))=G(c(t))$ for all $t\in [0,T]$. We will moreover show that
$U^{c}(t,x)=G(x)$ for $x\in[c(t),\infty)$ with $t\in [0,T]$. This can be derived using the martingale property as follows:
the Markov property of $X$ implies that
\begin{align}\label{th-5} \hs{3pc}
e^{-rs}U^{c}(&t\p s,X^x_s)-\int_0^{s} e^{-ru}H(X^x_u) I(X^x_u \ge c(t\p u))du=\;U^{c}(t,x)+N_s
\end{align}
where $(N_s)_{0\le s\le T-t}$ is a martingale under $\QQ$. On the other hand,
we know from \eqref{Tanaka-1} that
\begin{align} \label{th-6} \hs{3pc}
 e^{-rs}G(X^{x}_s)=\;G(x)+\int_0^s e^{-ru}H(X^{x}_u) I(X^x_u \ge K)du+M_s+\frac{1}{2}\int_0^s e^{-ru}d\ell^{K}_u (X^x)
 \end{align}
where $(M_s)_{0\le s\le T -t}$ is a continuous martingale under $\QQ$.

For $x\in[c(t),\infty)$ with $t\in[0,T]$ given and fixed, consider the stopping time
 \begin{align} \label{th-7} \hs{3pc}
\sigma_{c}=\inf\ \{\ 0\leq s\leq T\m t:c(t\p s)\ge X^x_{s}\ \}
 \end{align}
 under $\QQ$. Using that $U^{c}(t,c(t))=G(c(t))$ for all $t\in [0,T]$ and $U^{c}(T,x)=G(x)$ for all $x>0$, we see that
 $U^{c}(t\p\sigma_{c},X^x_{\sigma_{c}})=G(X^x_{\sigma_{c}})$. Hence,
 from \eqref{th-5} and  \eqref{th-6}, using the optional sampling theorem we find
 \begin{align}\label{th-8} \hs{3pc}
U^{c}(t,x)=\;&\EE e^{-r\sigma_{c}}U^{c}(t\p \sigma_{c},X^x_{\sigma_{c}})-\EE\int_0^{\sigma_{c}} e^{-ru}H(X^x_u) I(X^x_u\ge c(t\p u))du\\
=\;&\EE e^{-r\sigma_{c}}G(X^x_{\sigma_{c}})-\EE\int_0^{\sigma_{c}} e^{-ru}H(X^x_u)du=G(t,x)\nonumber
\end{align}
 as $X^x_u \in (c(t+ u),\infty)$ and $\ell^K_u (X^x)=0$ for all $u\in [0,\sigma_{c})$. This proves that $U^{c}(t,x)=G(x)$ for $x\in[c(t),\infty)$ with $t\in [0,T]$ as claimed.
 \vs{6pt}

 $(B.2)$ We show that $U^{c}(t,x)\le C^A(t,x)$ for all $(t,x)\in[0,T]\times(0,\infty)$.
 For this consider the stopping time
 \begin{align} \label{th-9} \hs{3pc}
\tau_{c}=\inf\ \{\ 0\leq s\leq T\m t:X^x_s\ge c(t\p s)\ \}
 \end{align}
 under $\QQ$ with $(t,x)\in[0,T]\times(0,\infty)$ given and fixed. The same arguments as those
 following \eqref{th-7} above show that $U^{c}(t\p\tau_{c},X^x_{\tau_{c}})=G(X^x_{\tau_{c}})$. Inserting $\tau_{c}$ instead of $s$ in \eqref{th-5} and using the optional sampling theorem, we get
 \begin{align}\label{th-10} \hs{3pc}
U^{c}(t,x)=\EE e^{-r\tau_{c}}U^{c}(t\p \tau_{c},X^x_{\tau_{c}})=
\EE e^{-r\tau_{c}}G(&X^x_{\tau_{c}})\le C^A(t,x)
\end{align}
 proving the claim.
 \vs{6pt}

  $(B.3)$ We show that $b\ge c$ on $[0,T]$. For this, suppose that there exists $t\in[0,T)$ such that $b(t)< c(t)$
 and choose a point $x\in [c(t),\infty)$ and consider the stopping time
  \begin{align} \label{th-11} \hs{3pc}
\sigma=\inf\ \{\ 0\leq s\leq T\m t:b(t\p s)\ge X^x_{s}\ \}
 \end{align}
 under $\QQ$. Inserting $\sigma$ instead of $s$ in \eqref{th-3} and \eqref{th-5} and using the optional sampling theorem, we obtain
 \begin{align} \label{th-12} \hs{3pc}
&\EE e^{-r\sigma}C^A(t\p \sigma,X^x_\sigma)=C^A(t,x) +\EE\int_0^\sigma e^{-ru}H(X^x_u)du\\
\label{th-13}&\EE e^{-r\sigma}U^{c}(t\p \sigma,X^x_\sigma)=U^{c}(t,x)+\EE\int_0^{\sigma}e^{-ru} H(X^x_u) I\big(X^x_u \ge c(t\p u))\big)du.
\end{align}
As $U^{c}\le C^A$ and $C^A(t,x)=U^{c}(t,x)=G(x)$ for
$x\in[c(t),\infty)$ with $t\in [0,T]$, it follows
from \eqref{th-12} and \eqref{th-13} that
\begin{align} \label{th-14} \hs{3pc}
\EE\int_0^{\sigma} e^{-ru}H(X^x_u) I\big(X^x_u\le c(t\p u)\big)du\ge0.
\end{align}
Due to the fact that $H$ is negative above $\max(K,x^*)$, we see by the continuity of $b$
and $c$ that \eqref{th-14} is not possible, so that we arrive at a contradiction. Hence, we can conclude that $b(t)\ge c(t)$ for all $t\in [0,T]$.
\vs{6pt}

$(B.4)$ We show that $c$ must be equal to $b$.
For this, let us assume that there exists $t\in[0,T)$ such that $c(t)<b(t)$. Choose an arbitrary point $x\in (c(t),b(t))$ and
consider the optimal stopping time $\tau^*$ from \eqref{problem-3}  under $\QQ$. Inserting $\tau^*$ instead of $s$ in \eqref{th-3} and \eqref{th-5}, and using the optional sampling theorem, gives
 \begin{align} \label{th-15} \hs{3pc}
&\EE e^{-r\tau^*}G(X^x_{\tau^*})=C^A(t,x)\\
\label{th-16}&\EE e^{-r\tau^*}G(X^x_{\tau^*})=U^{c}(t,x)+\EE\int_0^{{\tau^*}} e^{-ru}H(X^x_u) I\big(X^x_u \ge c(t\p u)\big)du
\end{align}
where we use that $C^A(t\p \tau^*,X^x_{\tau^*})=G(X^x_{\tau^*})=U^{c}(t\p \tau^*,X^x_{\tau^*})$ upon recalling that $c\le b$ and $U^{c}=G$ either above $c$ or at $T$. As $U^{c}\le C^A$, we have from
\eqref{th-15} and \eqref{th-16} that
 \begin{align} \label{th-17} \hs{3pc}
\EE \int_0^{{\tau^*}} e^{-ru}H(X^x_u) I\big(X^x_u \ge c(t\p u)\big)du\ge 0.
\end{align}
Due to the fact that $H$ is negative above $\max(K,x^*)$, we see from \eqref{th-17} by continuity of
$b$ and $c$ that such a point $(t,x)$ cannot exist. Thus, $c$ must be equal to $b$ and the proof of the theorem is complete.

\end{proof}

\begin{remark}
The integral equation \eqref{th-2} can be easily solved numerically via a backwards induction scheme based on a discretization of the integral with respect to time (for details see, e.g., Chapter 8 in Detemple (2006)). Note that in order to implement the algorithm, it is crucial to know the distribution of $Y_t$ and the value of $b(T)=\max(K,x^*)$.  See Figures 1 and 2 for illustrations  of the optimal exercise boundary
$b$ for models in Examples 2.1-2.6.
\end{remark}

\begin{remark}
Numerical computations using the EEP formula \eqref{th-1}
show that the American call price function $C^A$ fails to be convex with respect to $x$ under the $3/2$-model at $t=0$ (see Figure 3), unlike, e.g., in the geometric Brownian motion model. We note that the European call price function under the $3/2$-model is not convex either, which was also pointed out by Goard and Mazur (2013). In contrast, Figure 4 shows that, for the chosen set of parameters, the American call price function is convex in $x$ at $t=0$ under the $1/2$-model.
\end{remark}

\begin{figure}[t]
\begin{center}
\includegraphics[scale=0.7]{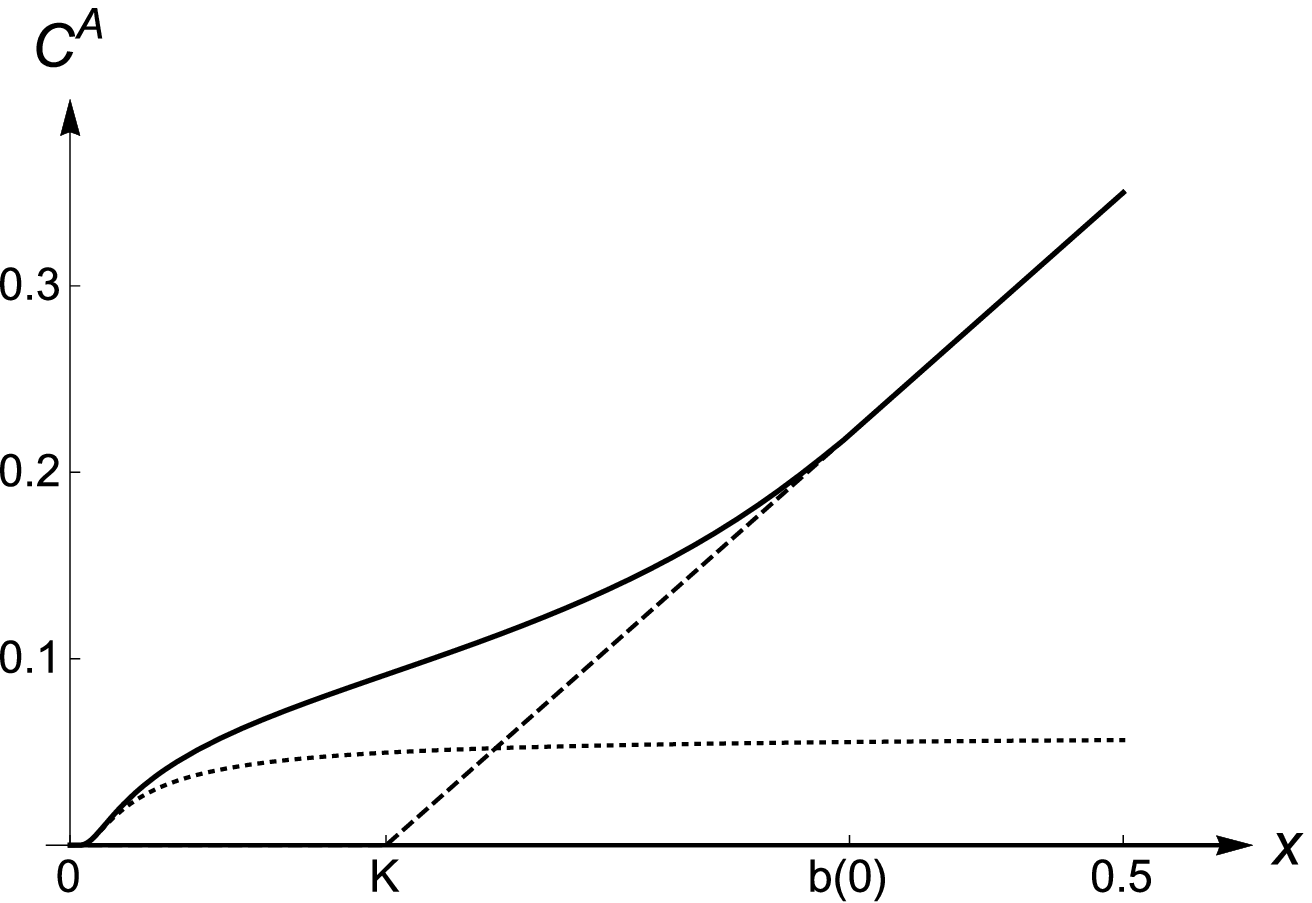}
\end{center}

{\par \leftskip=1.6cm \rightskip=1.6cm \small \ni \vs{-10pt}

\textbf{Figure 3.} This figure plots the price functions of the American $C^A(0,x)$ (solid) and the European $C^E(0,x)$ (dotted) call prices for the $3/2$-model against $x$ at $t=0$. The dashed line corresponds to the payoff function $(x\m K)^+$.
The graph shows that both functions fail to be convex with respect to $x$. The parameter set, as for Figure 1, is $T=1, K=0.15, r=0.05, \alpha=2.94, \beta=17.10, \kappa=2.05$.

\par} \vs{10pt}

\end{figure}

\begin{figure}[t]
\begin{center}
\includegraphics[scale=0.95]{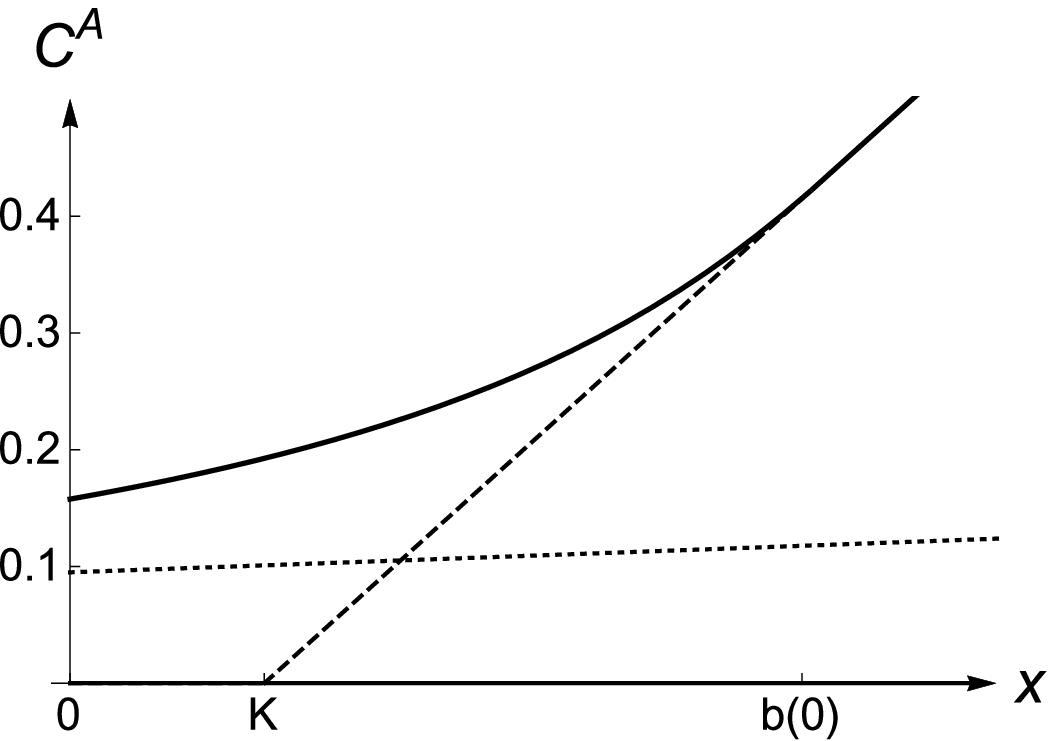}
\end{center}

{\par \leftskip=1.6cm \rightskip=1.6cm \small \ni \vs{-10pt}

\textbf{Figure 4.} This figure plots the price functions of the American $C^A(0,x)$ (solid) and the European $C^E(0,x)$ (dotted) call prices for the $1/2$-model against $x$ at $t=0$. The dashed line corresponds to the payoff function $(x\m K)^+$.
This graph shows that the American call price function is convex with respect to $x$ at $t=0$ for the parameter set $T=1, K=0.15, r=0.05$, $\alpha=2.94, \beta=17.10, \kappa=2.05$.

\par} \vs{10pt}

\end{figure}

 \section{The American VIX put option}

 In this section, we will briefly discuss the pricing problem for the American VIX put under model \eqref{vix-1} with $f$ of types $(A1)$ and $(A2)$
  \begin{equation} \label{put-1} \hs{6pc}
P^A(t,x)=\sup \limits_{0\le\tau\le T-t}\EE e^{-r\tau}\wt{G}(X^x_\tau)
 \end{equation}
 for $t\in[0,T)$ and $x>0$ where $\wt{G}(x)=(K\m x)^+$.
 The rational price of European VIX put  is given by
\begin{equation} \label{put-2} \hs{6pc}
P^E(t,x)=e^{-r(T-t)}\EE (K\m X^x_{T-t})^+
 \end{equation}
 for $t\in[0,T)$ and $x>0$. The latter can be computed in the same way as the European call in Section 2.
 The methodology for the American put option is very similar to the one for the call option, thus we omit an analysis and only state the main result.
As for the call option, here we impose the \textbf{Assumption R} on the function $h$.
\vs{6pt}

We define the function
\begin{align} \label{LL} \hs{5pc}
\wt{L}(u,x,z)=-\EE \big[e^{-ru}\wt{H}(X^x_u) I(X^x_u \le z)\big]
 \end{align}
 for $u\ge 0$ and $x,z>0$, which can be computed for $f$ of 3/2 type as follows
 \begin{align} \label{LL-1} \hs{5pc}
\wt{L}(u,x,z)=-e^{-ru}\int_{g(z)}^{\infty} \wt{H}(f(\wt{y}))\,q(\wt{y};u,g(x))\,d\wt{y}
 \end{align}
 and for $f$ of $1/2$ type as
  \begin{align} \label{LL-2} \hs{5pc}
\wt{L}(u,x,z)=-e^{-ru}\int_0^{g(z)} \wt{H}(f(\wt{y}))\,q(\wt{y};u,g(x))\,d\wt{y}.
 \end{align}
We now state the theorem on the rational price and optimal exercise boundary of the American VIX put. The proof is similar to the proof of Theorem 4.1.

 \begin{theorem}\label{th:2}
The optimal exercise strategy in \eqref{put-1} is given by
\begin{align} \label{OST-2} \hs{5pc}
&\wt{\tau}=\inf\ \{\ 0\leq s\leq T\m t:X^x_{s}\le \wt{b}(t\p s) \ \}
 \end{align}
 where the optimal exercise boundary $\wt{b}$ satisfies $0<\wt{b}(t)<\min(K,x^*)$ for $t\in[0,T)$ and $\wt{b}$ is increasing on $[0,T)$. The price function $P^A$ in \eqref{put-1} has the representation
\begin{align}\label{put-th-1} \hs{5pc}
P^A(t,x)=\;P^E(t,x) +\int_0^{T-t}\wt{L}(u,x,b(t\p u))du
\end{align}
for $t\in[0,T]$ and $x\in (0,\infty)$. The exercise boundary $\wt{b}$ in \eqref{put-1} can be characterized as the unique solution to the nonlinear integral equation
\begin{align}\label{put-th-2} \hs{5pc}
K-\wt{b}(t)=\;P^E(t,\wt{b}(t)) +\int_0^{T-t} \wt{L}(u,\wt{b}(t),\wt{b}(t\p u))du
\end{align}
for $t\in[0,T]$ in the class of continuous increasing functions $t\mapsto \wt{b}(t)$ with $\wt{b}(T)=\min(K,x^*)$.
\end{theorem}

\section{Pricing the American VIX call under the generalized mixture model}

In this section, we study the pricing of American VIX calls when the VIX is modelled as the sum of two processes: generalized 3/2- and 1/2-types.
In other words, the process $X$ is a function of a CIR process $Y$, where this function is the sum of functions of $(A1)$ and $(A2)$ types.
This can be seen as the generalization of the model introduced by Grasselli (2015), where the stochastic volatility is  $a/\sqrt{Y}\p b\sqrt{Y}$ and follows a $(2,0)$- mixture model in our terminology.
The process $Y$ represents the underlying factor for the optimal stopping problem.
In implementations of the model, this latent factor is calibrated. We will show that, under certain assumptions, there exists a pair of optimal exercise boundaries that
can be obtained as the unique solution to a system of coupled integral equations. The latter can be computed numerically by backward induction. We then provide the early exercise premium representation formula for the option price  which decomposes it into the sum of a  European part
 and an early exercise premium that depends on the pair of exercise boundaries.
\vs{6pt}

\subsection{The generalized mixture model}

1. Consider a mean-reverting square-root process (Feller or CIR process) under a risk neutral measure $\QQ$,%
\begin{align}\label{mix-1}\hs{6pc}
dY_t=(\beta\m\alpha Y_t)dt-\kappa\sqrt{Y_t}dB_t
\end{align}%
for $t>0$ where $B$ is a standard $\QQ$-Brownian motion started at $0$ and $\alpha
,\beta ,\kappa>0 $ are constant parameters such that $\beta\ge\kappa^2 /2$ (Feller
condition).

Now we take a function $f(y):=f_1(y)\p f_2(y)$ where $f_1$ is of $A1$-type and $f_2$ is of $A2$-type
and consider the VIX model
\begin{align}\label{mix-2}\hs{8pc}
X_t=f(Y_t)
\end{align}
for $t>0$.
Defining the processes $X_{1t}=f_1 (Y_t)$ (generalized 3/2-model) and $X_{2t}=f_2 (Y_t)$ (generalized 1/2-model), we then
obtain the alternative  characterization of $X$
\begin{align}\label{mix-3}\hs{8pc}
X_t=X_{1t}+X_{2t}
\end{align}
which means that $X$ is the mixture of generalized 3/2-type and 1/2-type of models. Throughout the section, we will mostly use \eqref{mix-2}.

It should be noted that in the mixture model, $X$ and $Y$ are not related to each other by a bijective function. Therefore the factor process $Y$ cannot be directly inferred from the observed
value of VIX, and there are two possible values for $Y$ for any given fixed $X$. However $Y$ can be easily calibrated from VIX futures prices, in particular, for short maturities (see Figure 6d).
Note also that $f$ converges to $+\infty$ as $Y$ goes to 0 or $+\infty$. We assume the following
\vs{2pt}

\textbf{Assumption M}: There exists $y_{\min}$ such that $f$  is strictly decreasing (increasing) for $y<y_{\min}$ $(y>y_{\min})$.

\textbf{Remark}. By differentiating the function $f$ one can see that \textbf{Assumption M} is equivalent to the following condition: $-f'_2(y)/f'_1(y) <1$ if and only if $y<y_{\min}$ for some $y_{\min}$.

Mixing the functions $f_1$ and $f_2$ from Examples 2.1-2.6, we can naturally consider the following models

\begin{example}
$((3/2,1/2)$-mixture model): Let $f(y) =a/y+by$ for positive
constants $a,b$. Then $f'(y) =-a/y^{2}+b, f''(y)=2a/y^{3}$. 
 The elasticity of variance is a non-linear function of the underlying factor.
\end{example}

\begin{example}
$((1 \p1/(2\nu), 1\m1/( 2\mu) )$-mixture model): Let $f(y)=a/y^{\nu }+b y^{\mu }$ where $\nu >0$, $\mu \in ( 0,1]$ and
$a, b$ are positive constants. Then $f'(y) =-a\nu /y^{\nu+1}+b\mu y^{\mu-1}, f''(y)=a\nu(\nu\p1)/y^{\nu+2}+b\mu(\mu\m1)y^{\mu-2}$.
The $(3/2,1/2)$-mixture model is obtained when $\nu=\mu=1$. The $(2,0)$-mixture model examined by Grasselli (2015) is obtained when $\nu=\mu=1/2$.
\end{example}

\begin{example}
$((1\p1/(2\nu _{j}), j=1,...,n, 1\m1/( 2\mu_i), i=1,...,m)$-mixture model): Let
$f(y)=\sum_{j}\omega _{j}/y^{\nu _{j}}+\sum_i \wh{\omega}_i y^{\mu_i}$ where $\nu _{j}>0, \omega
_{j}>0$ for $j=1,...,n$ and $\mu _{i}\in (0,1], \wh{\omega}_{i}>0$ for $i=1,...,n$ so that $f'(y)=-\sum_{j}\omega _{j}\nu_j /y^{\nu _{j}+1}+\sum_i \wh{\omega_i}\mu_i y^{\mu_i-1}$ and
$f''(y)=\sum_{j}\omega _{j}\nu_j(\nu_j +1) /y^{\nu _{j}+2}+\sum_i \wh{\omega}_i\mu_i(\mu_i \m 1) y^{\mu_i-2}$.
\end{example}

Examples 6.1-6.3 satisfy \textbf{Assumption M} as shown next for Example 6.3 which is the most general one.
Indeed, it is enough to show that the derivative $f'$ changes sign only once from negative to positive. Let us assume that $\mu_1<\mu_i$ for $i=2,...,m$,
then we rewrite $f'$ as
\begin{equation} \label{mix-assumption} \hs{3pc}
f'(y)=\frac{-\sum_{j}\omega _{j}\nu_j /y^{\nu _{j}+\mu_1}+ \wh{\omega}_1 \mu_1 +\sum_{i\ge2} \wh{\omega}_i\mu_i y^{\mu_i-\mu_1}}{y^{1-\mu_1}}
 \end{equation}
for $y>0$ and note that the numerator is strictly increasing and varies from $-\infty$ to $+\infty$ and the denominator is strictly positive.
Therefore $f'$ changes sign a single time and the proof of the initial claim is complete.
\vs{6pt}

2. Here we discuss the empirical relevance of the mixture model in Example 6.2.
 The Figures 5, 6a and 6b show possible slopes of implied volatility curves that can be generated by the model. Notably, it reproduces
 the positive skew and can fit the market data for VIX options well. Compared to the model in Example 2.2, it has two extra degrees of flexibility,
 coefficient $b$ and power $\mu$. In Figure 5, given a benchmark set of parameters $(\alpha, \beta, \kappa, r, T)$, we vary powers $(\nu,\mu)$
 and weights $(a,b)$. It can be seen (Figures 5c and 5d) that $a$ and $\nu$ are responsible for the parallel shifts of the volatility skew, and we note that
 low values of $\nu$ and relatively high values of $a$ can produce a smirk when moneyness is negative. Such a smirk is occasionally observed in the market. On the other hand, variations in $b$ and $\mu$ (Figures 5b and 5d) affect the slope of the skew, which is an important feature that helps to fit the market data. Overall, this analysis shows that by adding 1/2-type to 3/2-type of models we gain flexibility in capturing empirical features of the implied volatility curve.

 In Figures 6a and 6b, we explore the comparative statics of the volatility skew with respect to maturity $T$ and the diffusion term $\kappa$ of $Y$, respectively. It can be observed that implied volatility moves up when $T$ increases, which is consistent with empirical results documented, e.g., in Mencia and Sentana (2013). From Figure 6b, we note that relatively large or small values of $\kappa$ produce unrealistic levels of implied volatility. We also highlight that the effect of changes in $\alpha$ and $\beta$ on the volatility skew is negligible for the benchmark set of parameters.
 Finally, in Figures 6c and 6d, we provide the empirical curve of VIX futures prices  on particular days and the term structure given by the
 $((1 \p1/(2\nu), 1\m1/( 2\mu) )$-mixture model. In Figure 6d, we vary initial values of VIX, and then depending on the binary choice of $Y_0$ we can reproduce
 both upward and downward slopes for the term structure along with various forms of curvature. Moreover, as we mentioned, the changes in $\alpha$ and $\beta$ do not distort the volatility skew much so that they can be varied in order to fit the observed futures prices  .
\vs{6pt}

\begin{figure}[htb!]

\begin{center}
\includegraphics[width=.4\textwidth]{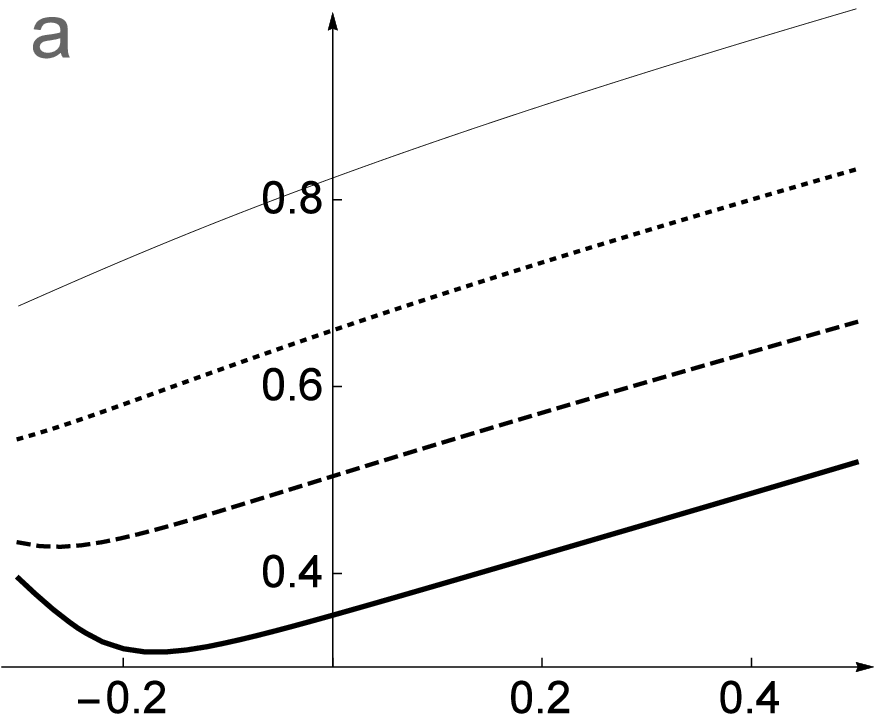}
\includegraphics[width=.4\textwidth]{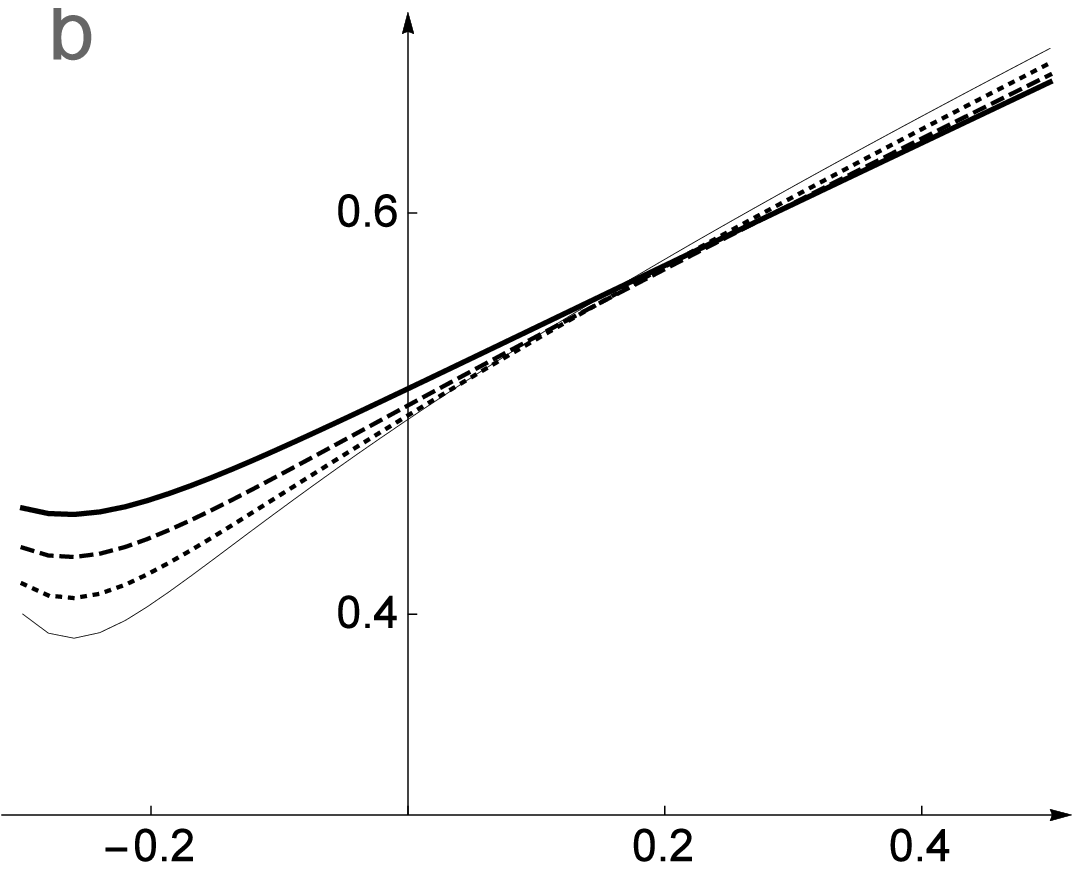}
\includegraphics[width=.4\textwidth]{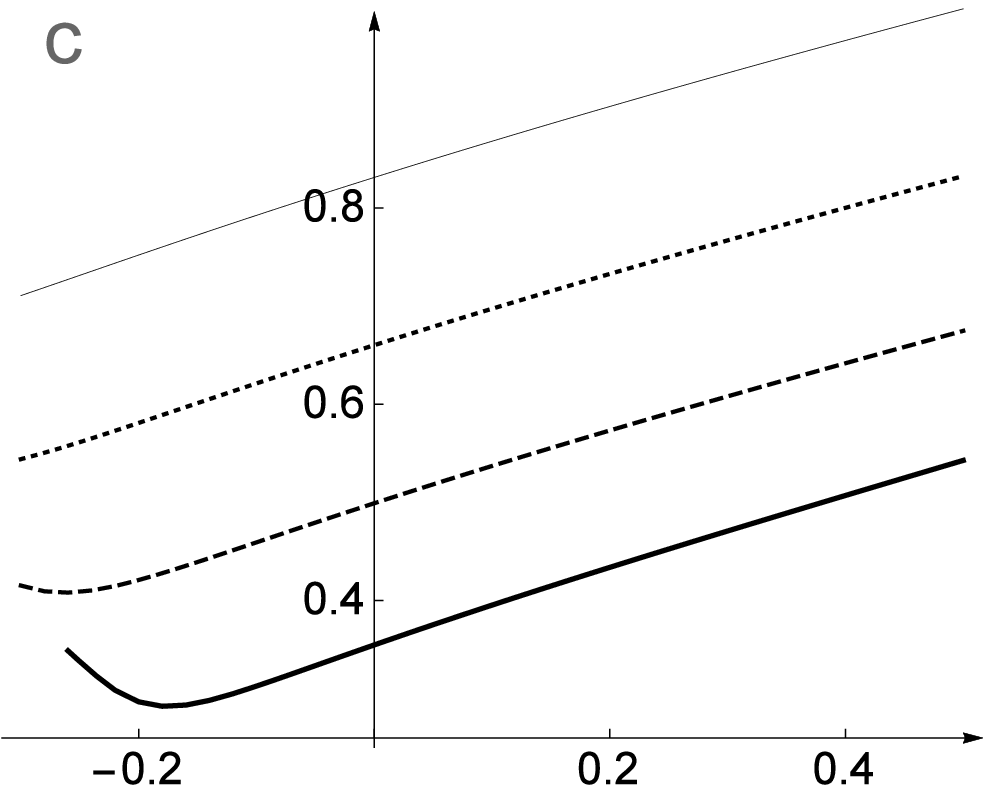}
\includegraphics[width=.4\textwidth]{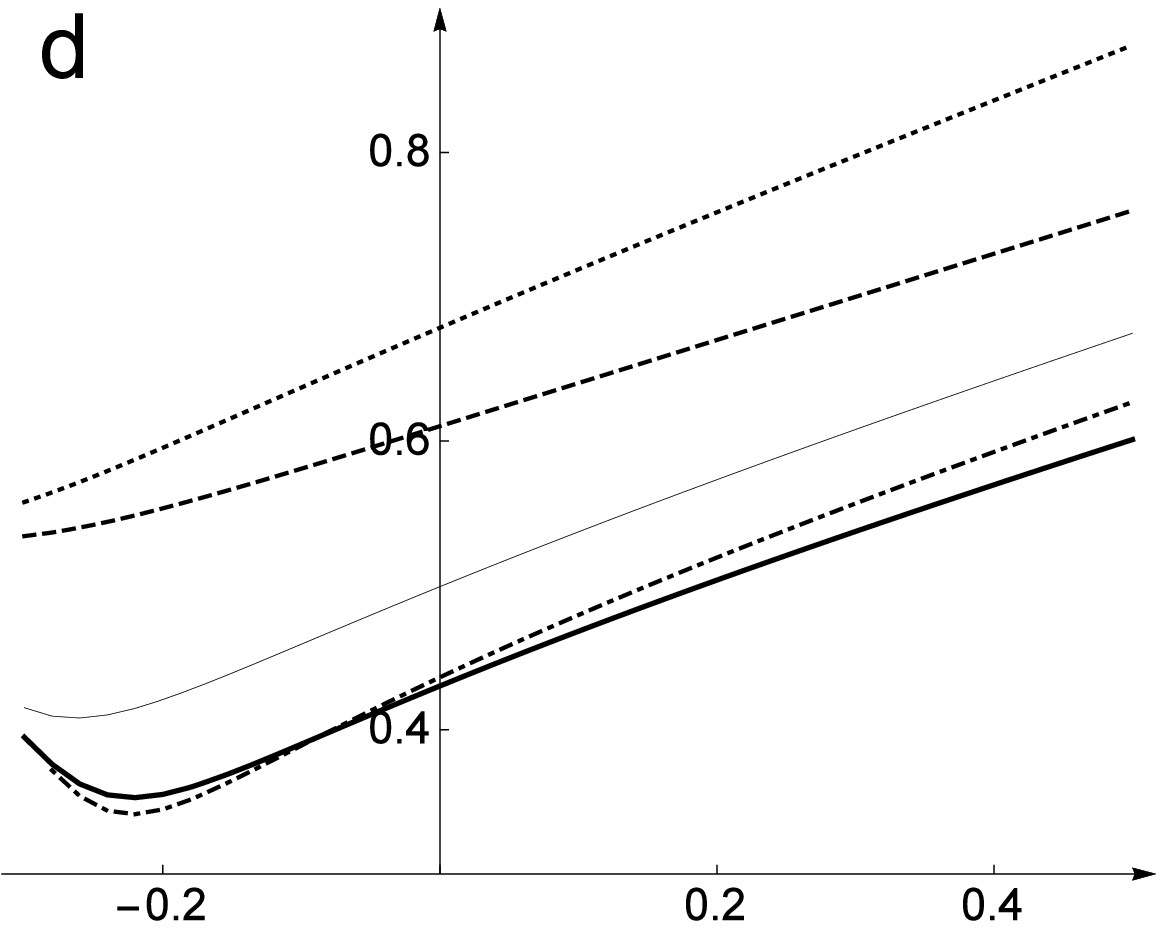}
\end{center}

{\par \leftskip=1.6cm \rightskip=1.6cm \small \ni \vs{-10pt}

\textbf{Figure 5.} Sensitivity analysis of implied volatility of VIX options under the $(1\p1/(2\nu), 1\m1/( 2\mu) )$-mixture model. Implied volatilities
for maturity $T$ are obtained by inverting the Black call price formula. The $y$-axis
represents the level of implied volatility, the $x$-axis records the moneyness $\log(K/F_T)$, where $K$ is the strike price and $F_T$ is the VIX futures price with maturity $T$.
 The benchmark set of parameters is: $\nu=0.75, \mu=1, a=0.1, b=0.02, r=0.01, \alpha=0.2, \kappa=0.7, \beta=0.1, T=2$ months, $X_0=0.137$, $Y_0=0.776$. \textbf{(a)} sensitivity w.r.t.  $\nu=\mu$: $\nu=\mu=0.5$ (thick), $\nu=\mu=0.75$ (dashed), $\nu=\mu=1$ (dotted), $\nu=\mu=1.25$ (thin).
 \textbf{(b)} sensitivity w.r.t.  $\mu$: $\nu=0.75, \mu=0.5$ (thick), $\nu=0.75, \mu=0.75$ (dashed), $\nu=0.75, \mu=1$ (dotted), $\nu=0.75, \mu=1.25$ (thin).
\textbf{(c)} sensitivity w.r.t.  $\nu$: $\nu=0.5, \mu=1$ (thick), $\nu=0.75, \mu=1$ (dashed), $\nu=1, \mu=1$ (dotted), $\nu=1.25, \mu=1$ (thin).
\textbf{(d)} sensitivity w.r.t. $a$ and $b$: $a=0.08, b=0.02$ (dotted), $a=0.1, b=0.01$ (dashed), $a=0.1, b=0.02$ (thin), $a=0.1, b=0.025$ (dot-dashed), $a=0.11, b=0.02$ (thick).

\par} \vs{10pt}

\end{figure}

\begin{figure}[htb!]

\begin{center}
\includegraphics[width=.4\textwidth]{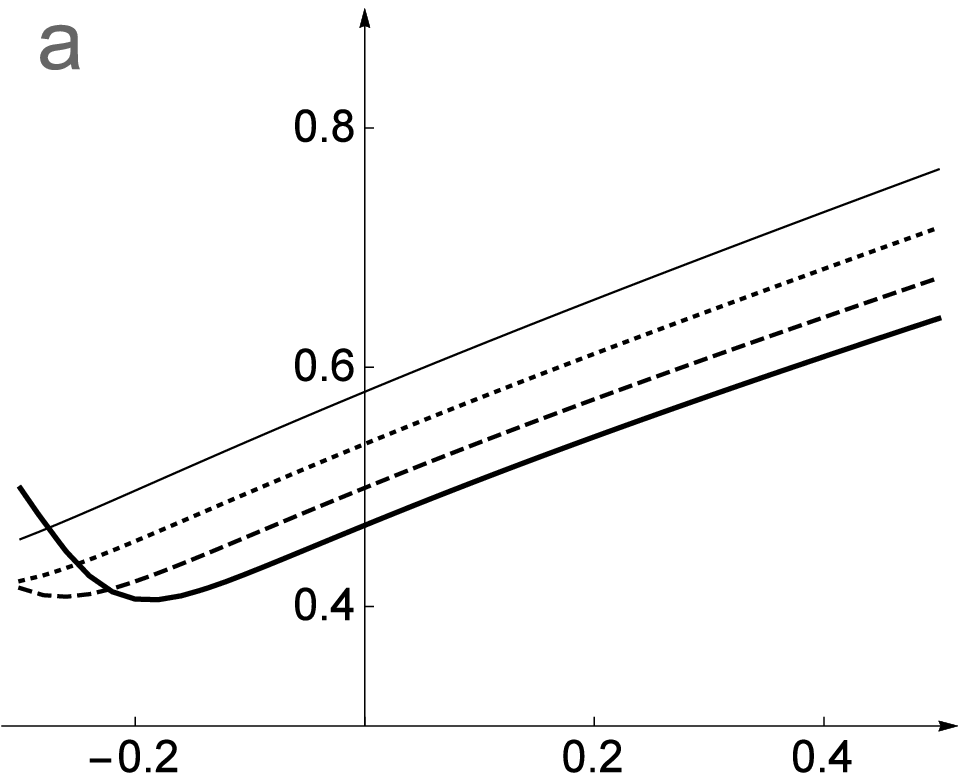}
\includegraphics[width=.4\textwidth]{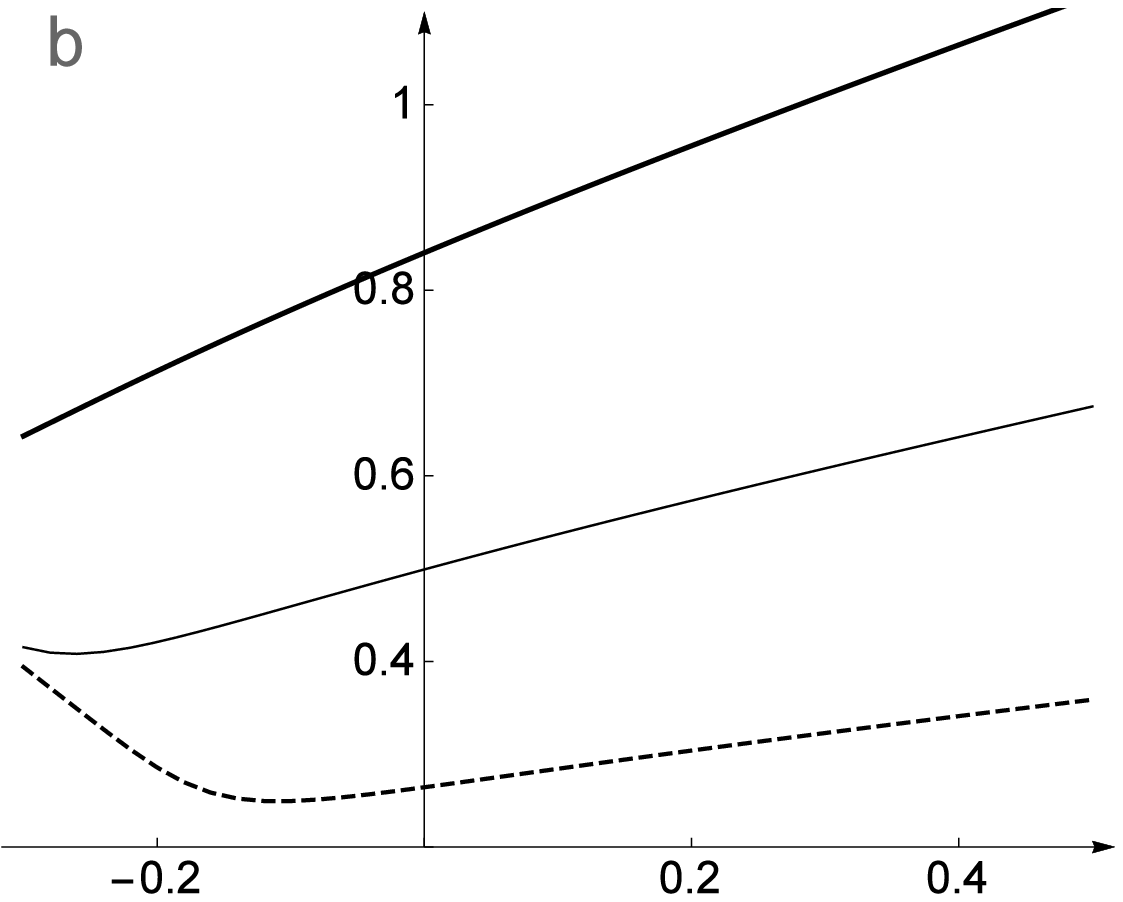}
\includegraphics[width=.4\textwidth]{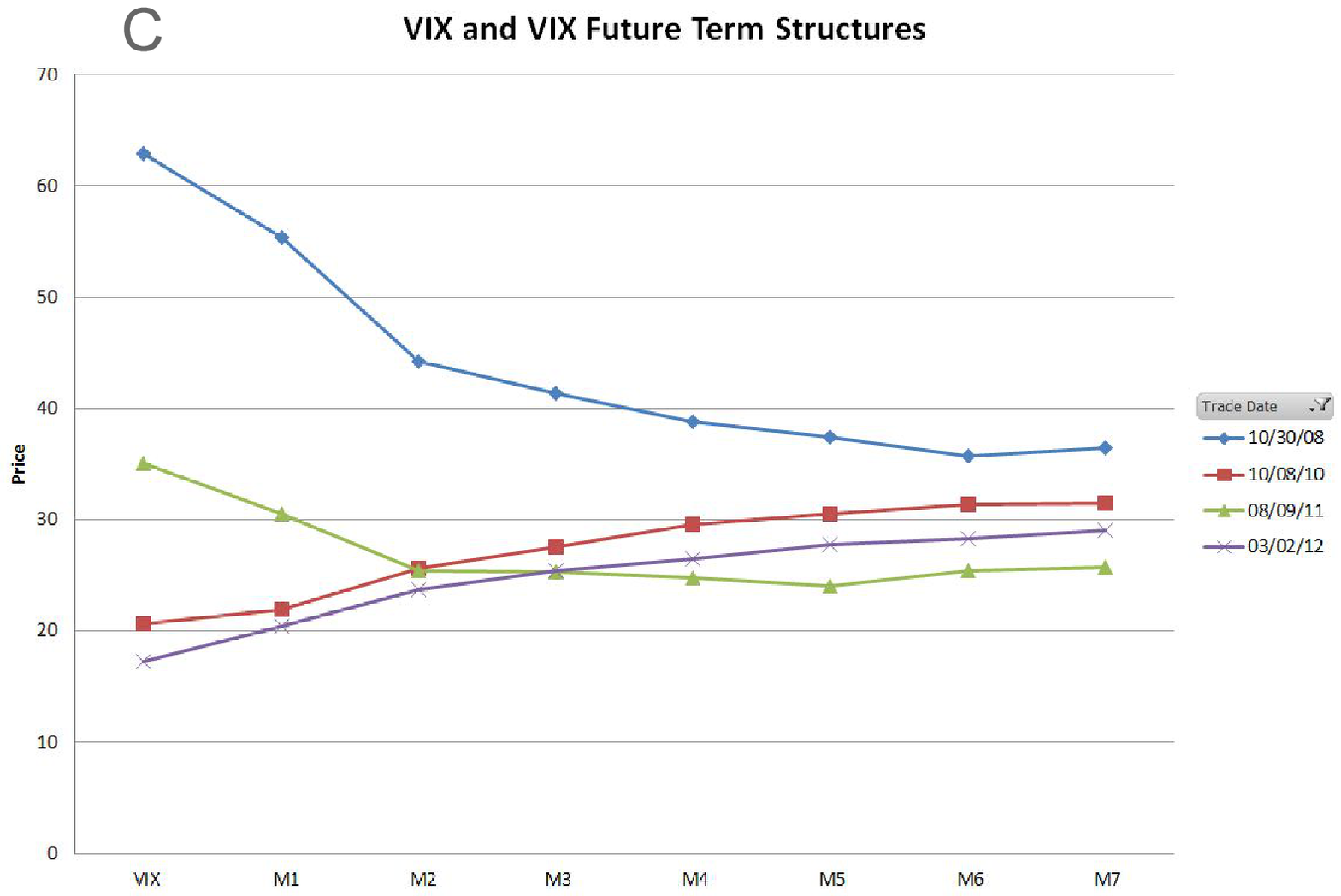}
\includegraphics[width=.4\textwidth]{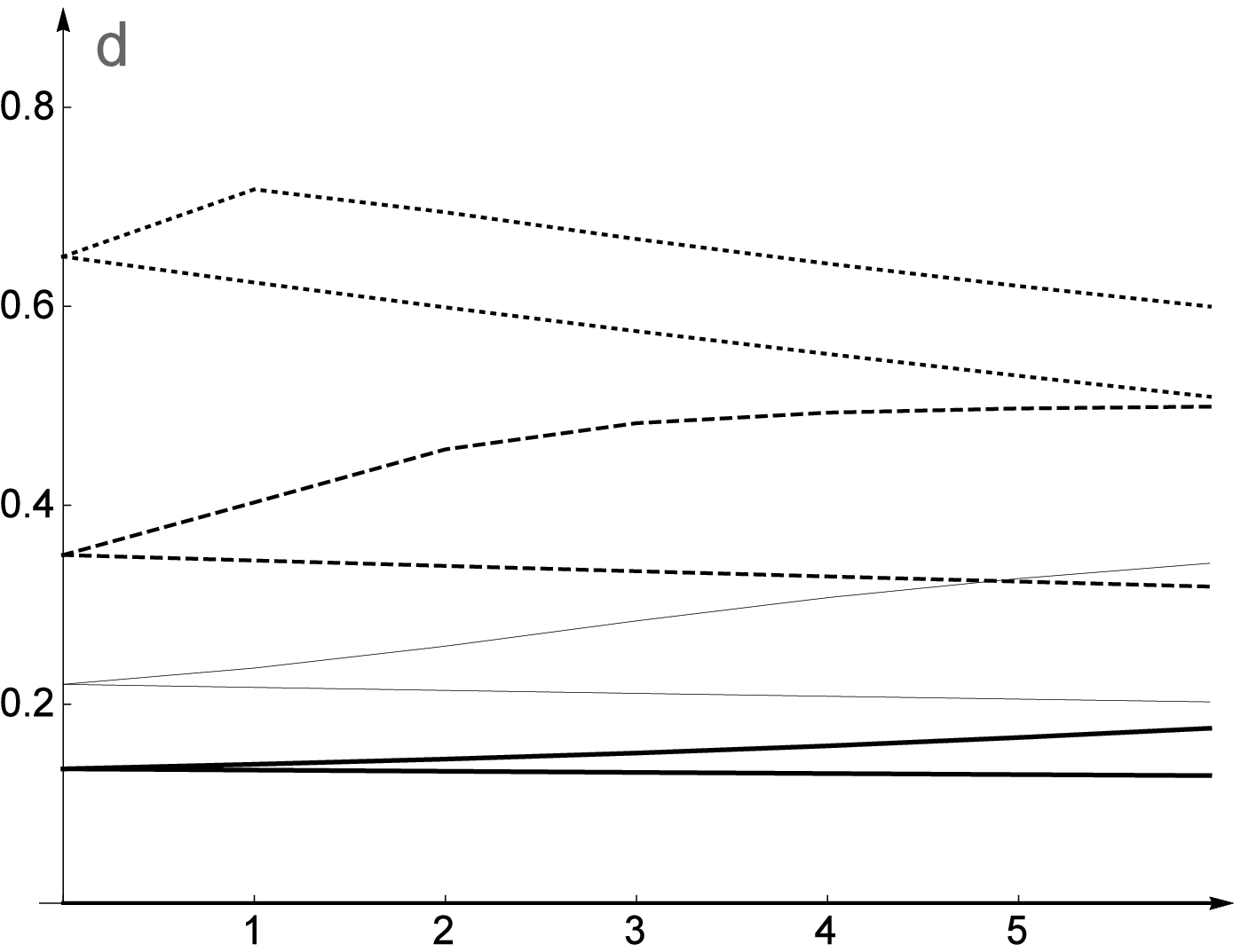}

\end{center}

{\par \leftskip=1.6cm \rightskip=1.6cm \small \ni \vs{-10pt}

\textbf{Figure 6.} Sensitivity analysis of implied volatility (panels \textbf{(a)} and \textbf{(b)}). Implied volatilities
for maturity $T$ are obtained by inverting the Black call price formula. The $x$-axis corresponds to moneyness $\log(K/F_T)$, where $K$ is the strike price and $F_T$ is the VIX futures price with maturity $T$. The benchmark set of parameters is: $\nu=0.75, \mu=1, a=0.1, b=0.02, r=0.01, \alpha=0.2, \kappa=0.7, \beta=0.1, T=2$ months, $X_0=0.137$, $Y_0=0.776$. \textbf{(a)}: sensitivity w.r.t.  maturity $T$: $T=1$ month (thick), $T=2$ months (dashed), $T=3$ months (dotted), $T=4$ months (thin).
 \textbf{(b)} sensitivity w.r.t.  $\kappa$: $\kappa=1$ (thick), $\kappa=0.7$ (thin), $\kappa=0.4$ (dashed).
Panels \textbf{(c)} and \textbf{(d)} show, respectively, the observed VIX futures term structure on particular days and the model VIX futures term structure for different initial values $X_0$ of VIX: $X_0=0.137$ (thick), $X_0=0.22$ (thin), $X_0=0.35$ (dashed),
$X_0=0.65$ (dotted); upper and lower curves correspond to two possible values of $Y_0$.

\par} \vs{10pt}

\end{figure}

3. Under the model \eqref{mix-1}-\eqref{mix-2},
 the rational price $C^A$ of the American VIX call at time $t=0$ is the value function of the following optimal stopping problem
\begin{equation} \label{mix-problem-0} \hs{6pc}
C^A=\sup \limits_{0\le\tau\le T}\EE e^{-r\tau} (f(Y_\tau)\m K)^+
 \end{equation}
where the supremum is taken over all stopping times $\tau$ of $Y$ and the expectation $\EE$ is taken under a risk neutral measure $\QQ$.

As the process $Y$ is time-homogeneous Markov, we will study the problem \eqref{mix-problem-0} in the Markovian setting and hence, we introduce dependence on time $t$ and the initial value of $Y$
 \begin{equation} \label{mix-problem} \hs{6pc}
C^A(t,y)=\sup \limits_{0\le\tau\le T-t}\EE e^{-r\tau}G(Y^y_\tau)
 \end{equation}
 for $t\in [0,T)$ and $y>0$, where $Y^y$ represents the process $Y$ started from $Y^y_0=y$ and the payoff function $G$ is given by
 \begin{equation} \label{mix-payoff} \hs{9pc}
 G(y):=(f(y)\m K)^+
 \end{equation}
for $y>0$.
\vs{6pt}

4. The rational price function of the European VIX call is
\begin{equation} \label{mix-eur} \hs{6pc}
C^E(t,y)=e^{-r(T-t)}\EE (f(Y^y_{T-t})\m K)^+
 \end{equation}
 for $t\in [0,T)$ and $y>0$. We note that given \textbf{Assumption M}, there are unique points $K_*\le K^*$ such that
 $f(y)\ge K$ when $y\le K_*$ or $y\ge K^*$.
 As the random variable $Y^y_t$ has non-central chi-squared density function
 $q(\wt{y};t,y)$,
one can compute $C^E$ numerically using
\begin{equation} \label{mix-eur-1} \hs{1pc}
C^E(t,y)=e^{-r(T-t)}\left[\int_{0}^{K_*}(f(\wt{y})-K)\,q(\wt{y};T\m t,y)\,d\wt{y}+
\int_{K^*}^{\infty}(f(\wt{y})-K)\,q(\wt{y};T\m t,y)\,d\wt{y}\right]
 \end{equation}
 for $t\in [0,T)$ and $y>0$.
 \vs{6pt}

 5. The VIX futures can be computed efficiently by straightforward numerical integration
\begin{equation} \label{mix-Fut-1} \hs{6pc}
F_{T}=\int_0^{\infty} f(\wt{y})\,q(\wt{y};T,y_0)\,d\wt{y}
 \end{equation}
 for $T>0$ (see Figure 6d). As in Section 2, we can approximate the futures price as follows
 \begin{equation} \label{mix-Fut-2} \hs{2pc}
F_{T}=\EE[f(Y_T)]\approx f(\EE[Y_T])+\sum_{k=2}^4 \frac{1}{k!}\EE (Y_T-\EE[Y_T])^k f^{(k)}(\EE[Y_T])
 \end{equation}
 where centered moments of $Y_T$ are known.
 \vs{6pt}

 \subsection{The free-boundary problem for the American VIX call}

 In this section, we reduce the problem \eqref{mix-problem} to a free-boundary problem which will be tackled using again the local time-space calculus (Peskir (2005a)).
The continuity of $G$ and standard arguments show that the continuation and exercise regions read
\begin{align} \label{mix-C} \hs{5pc}
&\cC= \{\, (t,y)\in[0,T)\! \times\! [0,\infty):C^A(t,y)>G(y)\, \} \\[3pt]
 \label{mix-D}&\cE= \{\, (t,y)\in[0,T)\! \times\! [0,\infty):C^A(t,y)=G(y)\, \}
 \end{align}
and the optimal stopping time in \eqref{mix-problem} is given by
\begin{align} \label{mix-OST} \hs{5pc}
\tau=\inf\ \{\ 0\leq s\leq T\m t:(t\p s,Y^y_{s})\in\cE\ \}.
 \end{align}

 The process \eqref{mix-1} also satisfies the conditions of Theorem 37 of Chapter V, Section 7  in Protter (1990)
so that
\begin{equation} \label{mix-flow} \hs{5pc}
\left[\EE \sup\limits_{0\leq u\leq T}\left(Y^x_u\m Y^y_u\right)^2\right]^{1/2}\le C_L \left|x-y\right|
 \end{equation}
for $x,y>0$ and some constant $C_L>0$. We will use this estimate for the proof of the smooth-fit property.
\vs{6pt}

1. First, we show that the price function $C^A$ is continuous on $[0,T)\times (0,\infty)$.
We have
\begin{align} \label{mix-cont-1} \hs{5pc}
 0\le&\; C^A(t,x)-C^A(t,y)\le\sup \limits_{0\leq\tau\leq T-t}\EE e^{-r\tau}\left(f(Y^x_\tau)\m f(Y^y_\tau)\right)^+\\
  \le&\EE \sup\limits_{0\leq u\leq T-t}\left(X^{f_1(x)}_{1u}\p X^{f_2(x)}_{2u}\m X^{f_1(y)}_{1u}\m X^{f_2(y)}_{2u}\right)^+ \nonumber\\
 \le&
 \EE \sup\limits_{0\leq u\leq T-t}\left(X^{f_1(x)}_{1u}\m X^{f_1(y)}_{1u}\right)^+ + \EE \sup\limits_{0\leq u\leq T-t}\left(X^{f_2(x)}_{2u}\m X^{f_2(y)}_{2u}\right)^+\nonumber
 \end{align}
for $x\ge y$ and $t\in [0,T)$, where we used that $\sup(f)-\sup(g)\leq\sup(f\m g)$,
$(x-K)^{+}-(y-K)^{+}\leq(x-y)^{+}$ for $x,y,K\in \R$, and the representation \eqref{mix-3}. Using the continuity of $f_1$ and $f_2$ and the same arguments for processes $X_1$ and $X_2$ as in paragraph 1 of Section 3, shows that  $y\mapsto C^A(t,y)$ is continuous uniformly over $t\in [0,T]$. The proof that $t\mapsto C^A(t,y)$ is continuous on $[0,T]$ for each $y\geq0$ fixed is also analogous to the one in paragraph 1 of Section 3 and thus we omit it.
Combining both facts establishes the continuity of $C^A$ on $[0,T)\times (0,\infty)$.
\vs{6pt}

2. Now we derive some initial insights into the structure of exercise region $\cE$.
\vs{2pt}

$(i)$ We first calculate the function $H(y)\!:=(\L_Y G \m rG)(y)$ for $y\in(0,\infty)$ (which is the instantaneous benefit
of waiting to exercise) where
\begin{align} \label{mix-H-0} \hs{5pc}
\L_Y =\left(\beta\m\alpha y\right)\frac{d}{dy}+\frac{\kappa^2 y}{2} \,  \frac{d^2}{dy^2}
\end{align}
is the infinitesimal generator of $Y$.
As $G(y)=(f(y)\m K)^+$, we have  that
\begin{align} \label{mix-H-1} \hs{5pc}
H(y)=h(y)I(y\le K_* \;\text{or}\; y\ge K^*)
\end{align}
for $y\in(0,\infty)$ where
\begin{align} \label{mix-H-2} \hs{5pc}
h(y)=\left(\beta\m\alpha y\right)f'(y)+\frac{\kappa^2 y}{2} f''(y)-rf(y)+rK
\end{align}
for $y>0$.
The following condition is imposed on the model
\vs{2pt}

\noindent \textbf{Assumption R'}: There exist $y_*<y^*$ such that
$H(y) \ge 0$ if and only if $\min(y_{*}, K_{*})\le y\le \max(y^{*},K^{*})$.

Numerical computations show that the models in Examples 6.1-6.3 satisfy this assumption for a wide range of parameters.
\vs{6pt}

$(ii)$  We now use the Ito-Tanaka's formula and the definition of $H$ to obtain
\begin{align} \label{mix-Tanaka-1} \hs{0pc}
 \EE e^{-r\tau}G(Y^{y}_\tau)=&\;G(y)+\EE \int_0^\tau e^{-rs}H(Y^{y}_s)ds\\
 &+\frac{1}{2}\EE \int_0^\tau e^{-rs}(-f'(K_*))d\ell^{K_*}_s (Y^y) +\frac{1}{2}\EE \int_0^\tau e^{-rs}f'(K^*)d\ell^{K^*}_s (Y^y)\nonumber
 \end{align}
for $y\in(0,\infty)$ and any stopping time $\tau$ of the process $Y$, where $(\ell^K_s(X))_{s\ge 0}$ is the local time process of $X$ at levels $K\in\{K_*, K^*\}$
\begin{align} \label{mix-Tanaka-2} \hs{3pc}
 \ell^{K}_s (X^x):=\QQ-\lim_{\eps \downarrow 0}\frac{1}{2\eps}\int_0^{s} I(K\m\eps<X^x_u<K\p\eps)\,d\left \langle X,X \right \rangle_u
 \end{align}
 and $d\ell^{K}_s (X^x)$ refers to the integration with respect to the continuous increasing function $s\mapsto \ell^{K}_s (X^x)$.

Equation \eqref{mix-Tanaka-1} and \textbf{Assumption R'} show that it is not optimal to exercise the call option when $\min(y_*,K_*)\le Y_t \le \max(y^*,K^*)$
 as $H(Y_t)\ge0$ in this region and thus both integral terms
on the right-hand side of \eqref{mix-Tanaka-1} are non-negative. This fact can be also explained in the particular case where $K_*\le Y_t\le K^*$ as follows:
if the option holder exercises between $K_*$ and $K^*$ the payoff is null, however there is a positive probability of receiving a strictly positive payoff in future.

Another implication of \eqref{mix-Tanaka-1} is that the exercise region is non-empty for all $t\in[0,T)$, as for small $y\downarrow 0$ and large $y\uparrow \infty$ the integrand $H$
is negative and the local time terms are zero, and thus due to the insufficient time to compensate for the negative $H$, it is optimal to stop at once.
\vs{6pt}

3. Next we prove further properties of the exercise region $\cE$ and define the optimal exercise boundaries.
\vs{2pt}

$(i)$ Using the same arguments as in Section 3, we can show that $\cE$ is right-connected.
\vs{2pt}

$(ii)$ Now let us take $t>0$ and $x>y>\max(K^*,y^*)$ such that $(t,y)\in \cE$.
Then, by right-connectedness of the exercise region, we have that $(s,y)\in \cE$ as well for any $s>t$. If we now run the process $(t,Y_t)$ from $(t,x)$, we cannot hit the level $\max(K^*,y^*)$ before exercise (as $x>y$), thus the local time terms in \eqref{mix-Tanaka-1} are 0 and the integrand $H$ is negative (by \textbf{Assumption R'}).
Therefore, it is optimal to exercise at $(t,x)$, which establishes up-connectedness of the exercise region $\cE$ when $y>\max(K^*,y^*)$.
Exploiting the same arguments, we show down-connectedness of the exercise region $\cE$ when $y<\min(K_*,y_*)$.
\vs{2pt}

$(iii)$ From $(i)$-$(ii)$ and paragraph 2$(ii)$ above, we can conclude that
there exist a pair of optimal exercise boundaries $b_*:[0,T]\rightarrow (0,\infty)$
and $b^*:[0,T]\rightarrow (0,\infty)$ such that
\begin{align} \label{mix-OST-2} \hs{4pc}
&\tau=\inf\ \{\ 0\leq s\leq T\m t:Y^y_s \le b_* (t\p s) \; \text{or}\; Y^y_{s}\ge b^*(t\p s) \ \}
 \end{align}
is optimal in \eqref{mix-problem} and $0<b_*(t)<\min(K_*,y_*)<\max(K^*,y^*)<b^*(t)<\infty$ for $t\in[0,T)$. Moreover, $b_*$ is increasing and $b^*$ is decreasing on $[0,T)$.
\vs{6pt}

4. Now we prove that the smooth-fit condition along the boundaries $b_*$ and $b^*$ holds
 \begin{align}\label{mix-SF}\hs{4pc}
&C^A_y (t, b_*(t)+)=C^A_y(t,b_*(t)-)=G'(b_*(t))=f'(b_*(t))\\
\label{mix-SFa}&C^A_y (t, b^*(t)-)=C^A_y(t,b^*(t)+)=G'(b^*(t))=f'(b^*(t))
\end{align}
for all $t\in[0,T)$. We will only prove \eqref{mix-SFa} below, as the proof for the lower boundary $b_*$ is similar and can be omitted.
\vs{2pt}

$(i)$ First, let us fix a point $(t,y)\in [0,T)\times(0,\infty)$ lying on the boundary $b^*$ so that $y=b^*(t)$.
Then, we have
\begin{align} \label{mix-SF-1} \hs{4pc}
\frac{C^A(t,y)-C^A(t,y\m\eps)}{\eps}&\le \frac{G(y)-G(y\m\eps)}{\eps}
\end{align}
and taking the limit as $\eps\downarrow 0$, we get
\begin{align} \label{mix-SF-2} \hs{4pc}
\limsup_{\eps\downarrow 0} \frac{C^A(t,y)-C^A(t,y\m\eps)}{\eps}\le G'(y)=f'(y).
\end{align}

$(ii)$ To prove the reverse inequality, we set $\tau_\eps=\tau_\eps(t,y\m\eps)$ as an optimal stopping time for $C^A(t,y\m\eps)$.
Using that $Y$ is a regular diffusion  and $t\mapsto b^*(t)$ is decreasing, we have that
$\tau_\eps\to0$ as $\eps\to0$ $\QQ$-a.s. By the comparison theorem for solutions
of SDEs and noting that%
\begin{align}\label{mix-SF-3a}\hs{0pc}
G( Y_{\tau _{\eps }}^{y}&) \m G( Y_{\tau _{\eps}}^{y-\eps })  \\
=&\left( f(Y_{\tau _{\eps }}^{y})\m f(Y_{\tau_{\eps }}^{y-\eps})\right) I\left(f(Y_{\tau _{\eps }}^{y-\eps})\ge K\right)+(f(Y_{\tau _{\eps }}^{y})\m K)
I\left(f(Y_{\tau _{\eps }}^{y})\geq K\geq f(Y_{\tau _{\eps }}^{y-\eps})\right) \nonumber\\
\geq &\left( f(Y_{\tau _{\eps }}^{y})\m f(Y_{\tau_{\eps }}^{y-\eps})\right) I\left(f(Y_{\tau _{\eps }}^{y-\eps})\ge K\right)
\nonumber
\end{align}%
we obtain
 \begin{align}\label{mix-SF-3}\hs{0pc}
\frac{1}{\eps}\Big(&C^A(t,y)-C^A(t,y\m\eps)\Big)\\
&\ge \frac{1}{\eps}
\EE \left[e^{-r\tau_{\eps}}\left( f(Y_{\tau _{\eps }}^{y})\m f(Y_{\tau_{\eps }}^{y-\eps})\right) I\left(f(Y_{\tau _{\eps }}^{y-\eps})\ge K\right)\right]\nonumber\\
&=\frac{1}{\eps}\EE \left[e^{-r\tau_{\eps}}\left( f(Y_{\tau _{\eps }}^{y})\m f(Y_{\tau_{\eps }}^{y-\eps})\right) \right]
-\frac{1}{\eps}\EE \left[e^{-r\tau_{\eps}}\left( f(Y_{\tau _{\eps }}^{y})\m f(Y_{\tau_{\eps }}^{y-\eps})\right) I\left(f(Y_{\tau _{\eps }}^{y-\eps})\le K\right)\right]\nonumber.
\end{align}
 Then the second term on the right-hand side of
\eqref{mix-SF-3} goes to 0 as $\eps\to0$ as
 \begin{align}\label{mix-SF-4}\hs{2pc}
0&\le \frac{1}{\eps}\EE \left[e^{-r\tau_{\eps}}\left( f(Y_{\tau _{\eps }}^{y})\m f(Y_{\tau_{\eps }}^{y-\eps})\right)
I\left(f(Y_{\tau _{\eps }}^{y-\eps})\le K\right)\right]\\
&\le \frac{1}{\eps}\left(\EE \left( f(Y_{\tau _{\eps }}^{y})\m f(Y_{\tau_{\eps }}^{y-\eps})\right)^2\right)^{1/2}
\left(\QQ(f(Y_{\tau _{\eps }}^{y-\eps})\le K)\right)^{1/2}\nonumber\\
&= \frac{1}{\eps}\left(\EE \left( f'(\xi)(Y_{\tau _{\eps }}^{y}\m Y_{\tau_{\eps }}^{y-\eps})\right)^2\right)^{1/2}
\left(\QQ(f(Y_{\tau _{\eps }}^{y-\eps})\le K)\right)^{1/2}\nonumber\\
&\le \frac{1}{\eps}C_{f'}\left(\EE \sup\limits_{0\leq u\leq T-t}\left( Y_{u}^{y}\m Y_{u}^{y-\eps}
\right)^2\right)^{1/2} \left(\QQ(f(Y_{\tau _{\eps }}^{y-\eps})\le K)\right)^{1/2}\nonumber\\
&\le C_{f'} C_L \left(\QQ(f(Y_{\tau _{\eps }}^{y-\eps})\le K)\right)^{1/2}\rightarrow 0\nonumber
\end{align}
where we used Holder inequality, the mean value theorem with $\xi\in[Y_{\tau _{\eps }}^{y-\eps},Y_{\tau _{\eps }}^{y}]$, the facts that $|f'(y)|\le C_{f'}$ for some constant $C_{f'}>0$ and any $y\ge b_* (0)>0$, that $\xi\ge Y_{\tau _{\eps }}^{y-\eps}\ge b_*(t\p\tau_{\eps})>b_*(0)$, the inequality \eqref{mix-flow} and that the latter probability goes to zero because $y>K^*$.
 Now, we turn to the first term on the right-hand side of \eqref{mix-SF-3}. Using Ito's formula we have
 \begin{align}\label{mix-SF-5}\hs{0pc}
\frac{1}{\eps}\EE \left[e^{-r\tau_\eps}\left( f(Y_{\tau _{\eps }}^{y})\m f(Y_{\tau _{\eps }}^{y-\eps})\right)\right]=\frac{f(y)\m f(y\m\eps)}{\eps}+\frac{1}{\eps}
\EE \left[\int_0^{\tau_\eps} e^{-rs}\left(\omega(Y^y_s) \m \omega(Y^{y-\eps}_s)\right)ds\right]
\end{align}
where $\omega(y):=\big(\beta\m\alpha y\big)f'(y)\p\frac{1}{2}\kappa^2 yf''(y)\m r f(y)$ for $y>0$. We show that the second term of \eqref{mix-SF-5} goes to 0 as $\eps\to0$
\begin{align}\label{mix-SF-6}\hs{2pc}
0&\le \frac{1}{\eps}
\EE \left|\int_0^{\tau_\eps} e^{-rs}(\omega(Y^y_s) \m \omega(Y^{y-\eps}_s))ds
\right|\le \frac{1}{\eps}
\left[\EE \int_0^{\tau_\eps} e^{-rs}|\omega'(\xi_s)|(Y^y_s \m Y^{y-\eps}_s)ds\right]\\
&\le \frac{1}{\eps}
C_{\omega'} \EE \left[{\tau_\eps} \sup\limits_{0\leq u\leq T-t}\left( Y_{u}^{y}\m Y_{u}^{y-\eps
}\right)\right]\le \frac{1}{\eps} C_{\omega'} \left(\EE {\tau^2_\eps}\right)^{1/2} \left(\EE \sup\limits_{0\leq u\leq T-t}\left( Y_{u}^{y}\m Y_{u}^{y-\eps
}\right)^2 \right)^{1/2}\nonumber\\
&\le \frac{1}{\eps} C_{\omega'} C_L\, \eps \left(\EE {\tau^2_\eps}\right)^{1/2}=C_{\omega'} C_L \left(\EE {\tau^2_\eps}\right)^{1/2} \rightarrow 0\nonumber
\end{align}
where we used the mean value theorem with $\xi_s\in[Y^{y-\eps}_s,Y^{y}_s]$, the facts that $|\omega'(y)|\le C_{\omega'}$ for some
$C_{\omega'}>0$ and all $y\ge b_*(0)>0$, that $\xi_s\ge Y^{y-\eps}_s\ge b_*(t)>b_*(0)$ for $s\in[0,\tau_\eps]$, Holder inequality, the inequality \eqref{mix-flow} and that $\EE \tau^2_\eps \to0 $ as
$\eps\to0$ by the dominated convergence theorem.

 Thus, using \eqref{mix-SF-3}-\eqref{mix-SF-6} and taking the limits as $\eps\to0$ we have that
\begin{align} \label{mix-SF-7} \hs{2pc}
\liminf_{\eps\downarrow 0} \frac{C^A(t,y)-C^A(t,y\m\eps)}{\eps}\ge G'(y)=f'(y)
\end{align}
for $t\in[0,T)$. Thus, combining \eqref{mix-SF-2} and \eqref{mix-SF-7} we obtain \eqref{mix-SFa}.
\vs{6pt}

5. Using similar arguments as in paragraph 5 of Section 3, we can prove that the boundaries $b_*$ and $b^*$ are continuous on $[0,T]$ and that $b_*(T-)=\min(K_*,y_*)$
and $b^*(T-)=\max(K^*,y^*)$.
\vs{6pt}

6. The facts proved in paragraphs 1-5 above and  standard arguments based on the strong Markov property (see, e.g., Peskir and Shiryaev (2006)) lead to the following free-boundary problem for the value function $C^A$ and unknown boundaries $b_*$ and $b^*$
\begin{align} \label{mix-PDE} \hs{5pc}
&C^A_t \p\L_Y C^A \m rC^A=0 &\hs{-30pt}\text{in}\;  \cC\\
\label{mix-IS1}&C^A(t,b_*(t))=G(b_*(t))=f(b_*(t))-K &\hs{-30pt}\text{for}\; t\in[0,T)\\
\label{mix-IS2}&C^A(t,b^*(t))=G(b^*(t))=f(b^*(t))-K &\hs{-30pt}\text{for}\; t\in[0,T)\\
\label{mix-SP1}&C^A_y (t,b_*(t))=G'(b_*(t))=f'(b_*(t)) &\hs{-30pt}\text{for}\; t\in[0,T) \\
\label{mix-SP2}&C^A_y (t,b^*(t))=G'(b^*(t))=f'(b^*(t)) &\hs{-30pt}\text{for}\; t\in[0,T) \\
\label{mix-FBP1}&C^A(t,y)>G(y) &\hs{-30pt}\text{in}\; \cC\\
\label{mix-FBP2}&C^A(t,y)=G(y) &\hs{-30pt}\text{in}\; \cE
\end{align}
where the continuation set $\cC$ and the exercise set $\cE$ are given by
\begin{align} \label{mix-C-1} \hs{5pc}
&\cC= \{\, (t,y)\in[0,T)\! \times\! (0,\infty):b_*(t)<y<b^*(t)\, \} \\[3pt]
 \label{mix-D-1}&\cE= \{\, (t,y)\in[0,T)\! \times\! (0,\infty):y\le b_*(t)\,\text{or}\,y\ge b^*(t)\, \}.
 \end{align}
The following properties of $C^A$, $b_*$ and $b^*$ were also verified above
\begin{align} \label{mix-Prop-1} \hs{3pc}
&C^A\;\text{is continuous on}\; [0,T]\times(0,\infty)\\
\label{mix-Prop-2}&C^A\;\text{is}\; C^{1,2}\;\text{on}\; \cC\\
\label{mix-Prop-4}&t\mapsto C^A(t,y)\;\text{is decreasing on $[0,T]$ for each $y\in [0,\infty)$}\\
\label{mix-Prop-5}&t\mapsto b_*(t)\;\text{is increasing and continuous on $[0,T]$ with}\; b_*(T-)=\min(K_*,x_*)\\
\label{mix-Prop-6}&t\mapsto b^*(t)\;\text{is decreasing and continuous on $[0,T]$ with}\; b^*(T-)=\max(K^*,x^*).
\end{align}

\vs{6pt}

7. We recall that we already showed how to compute the European VIX call price in Section 6.1 above. Now define the function
\begin{align} \label{mix-L} \hs{5pc}
L(u,y,z_1,z_2)=-\EE \big[e^{-ru}H(Y^y_u) I(Y^y_u \le z_1\,\text{or}\,Y^y_u\ge z_2)\big]
 \end{align}
 for $u\ge 0$ and $y,z_1,z_2>0$.
  Using that
the random variable $Y^y_t$ has non-central chi-squared density function
 $q(\wt{y};t,y)$, we have
 \begin{align} \label{mix-L-1} \hs{3pc}
L(u,y,z_1,z_2)=-e^{-ru}\int_0^{z_1} H(\wt{y})\,q(\wt{y};u,y)\,d\wt{y}-e^{-ru}\int_{z_2}^{\infty} H(\wt{y})\,q(\wt{y};u,y)\,d\wt{y}
 \end{align}
 for $u\ge 0$ and $y,z_1,z_2>0$.
  \vs{+6pt}

\begin{theorem}\label{th:3}
The price function $C^A$ in \eqref{mix-problem} has the representation
\begin{align}\label{mix-th-1} \hs{3pc}
C^A(t,y)=\;C^E(t,y) +\int_0^{T-t}L(u,y,b_*(t\p u),b^*(t\p u))du
\end{align}
for $t\in[0,T]$ and $y\in (0,\infty)$. The optimal exercise boundaries $b_*$ and $b^*$ in \eqref{mix-problem} can be characterized as the unique solution to the coupled nonlinear integral equations of Volterra type
\begin{align}\label{mix-th-2} \hs{3pc}
&f(b_*(t))-K=\;C^E(t,b_*(t)) +\int_0^{T-t} L(u,b_*(t),b_*(t\p u),b^*(t\p u))du\\
\label{mix-th-3}
&f(b^*(t))-K=\;C^E(t,b^*(t)) +\int_0^{T-t} L(u,b^*(t),b_*(t\p u),b^*(t\p u))du
\end{align}
for $t\in[0,T]$, in the class of continuous functions $b_*(t)$ and $b^*(t)$ with $b_*(T)=\min(K_*,y_*)$ and $b^*(T)=\max(K^*,y^*)$ (See Figures 7 and 8).
\end{theorem}

\begin{figure}[t]
\begin{center}
\includegraphics[scale=0.75]{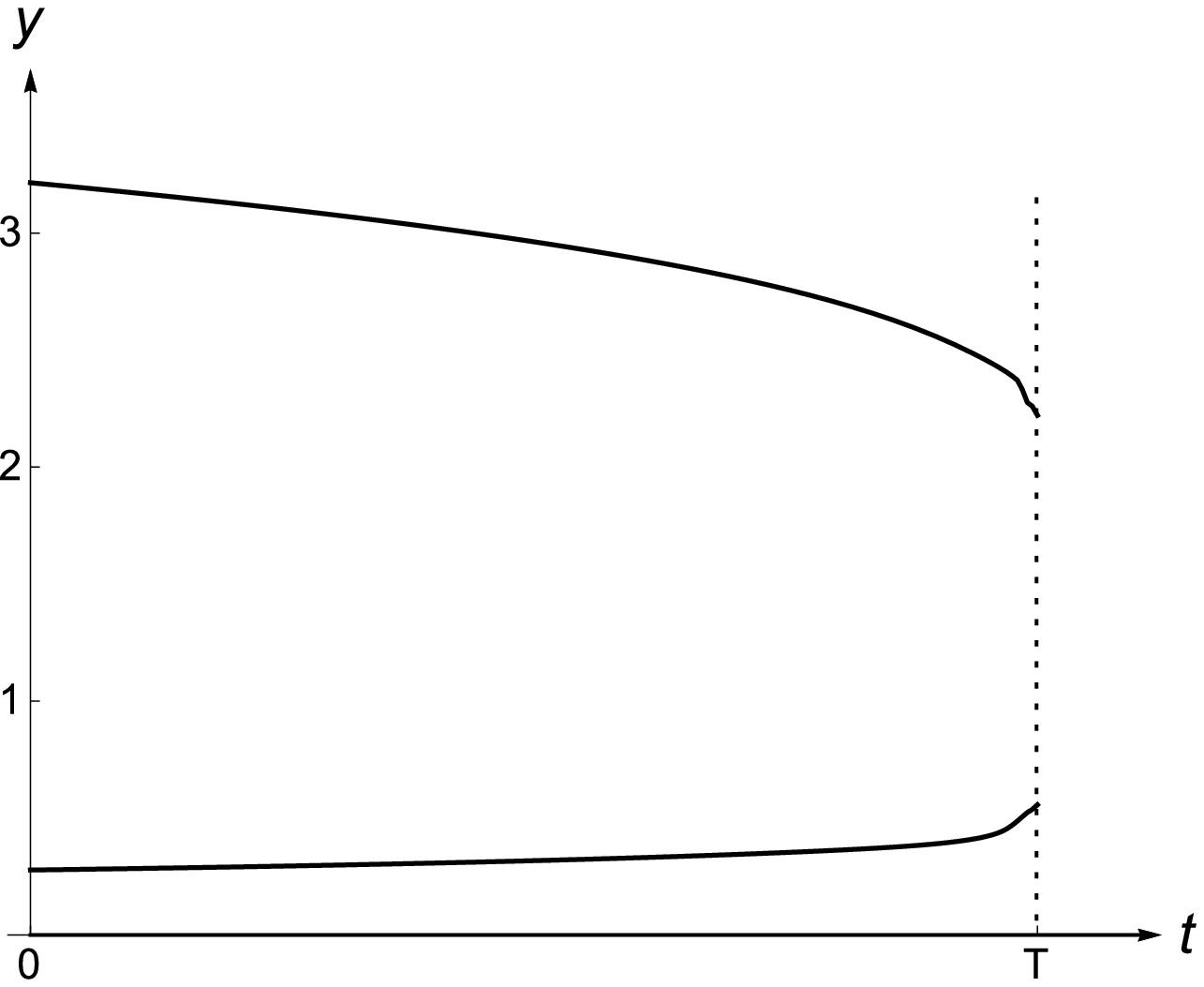}
\end{center}

{\par \leftskip=1.6cm \rightskip=1.6cm \small \ni \vs{-10pt}

\textbf{Figure 7.} This figure plots the optimal exercise boundaries $b_*$ (lower) and $b^*$ (upper) for the process $Y$ in the
$(3/2,1/2)$-mixture model.
The parameter set is $T=1$ year, $\alpha=\kappa=1, \beta=2, r=0.05, K = 0.15, a=b=0.07$.

\par} \vs{10pt}

\end{figure}

\begin{figure}[t]
\begin{center}
\includegraphics[scale=0.7]{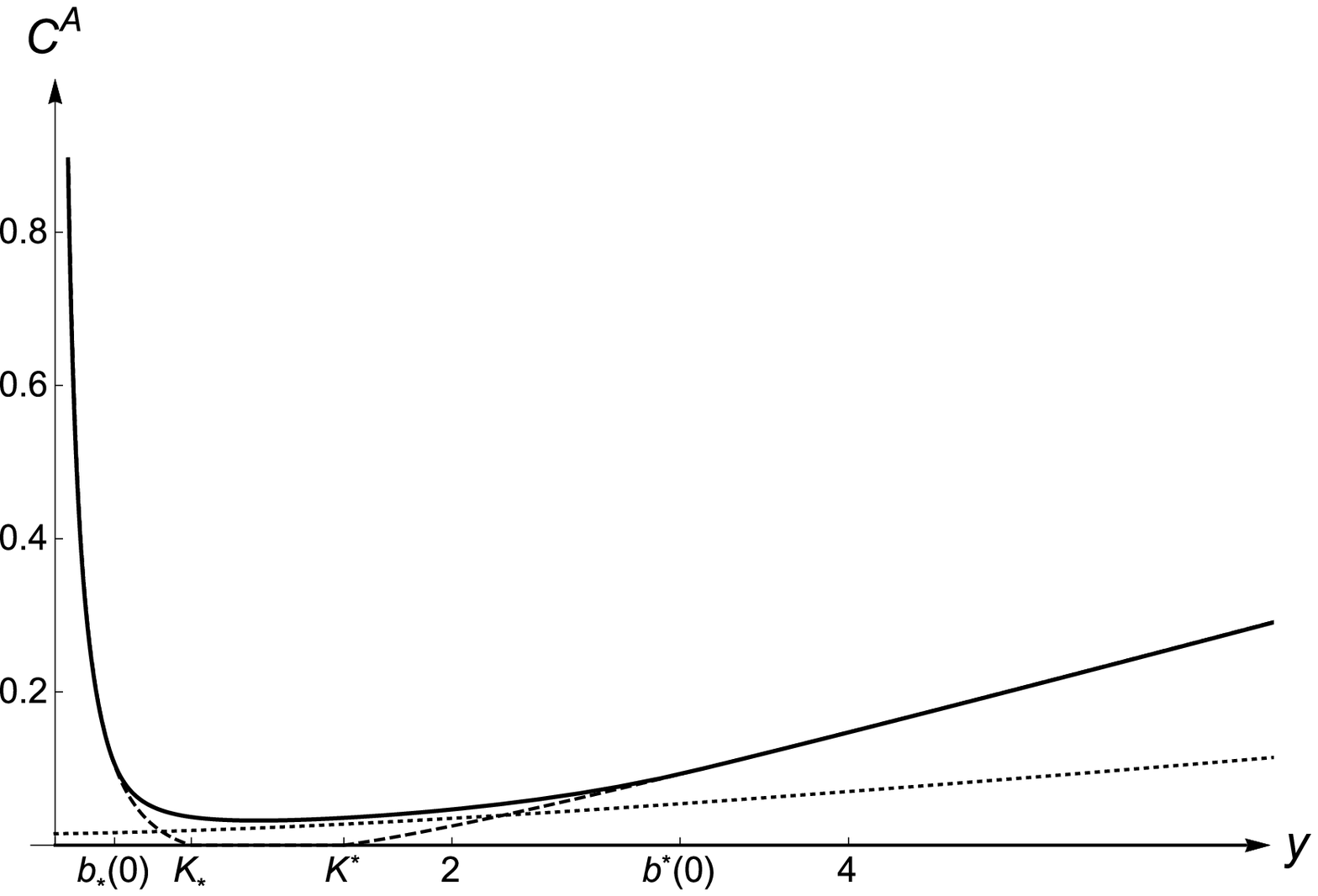}
\end{center}

{\par \leftskip=1.6cm \rightskip=1.6cm \small \ni \vs{-10pt}

\textbf{Figure 8.} This figure plots the price functions of the American $C^A(0,y)$ (solid) and the European $C^E(0,y)$ (dotted) call price functions for the $(3/2,1/2)$-mixture model against $y$ at $t=0$. The dashed line corresponds to the payoff function $(f(y)\m K)^+$.
The parameter set, as for Figure 5, is $T=1$ year, $\alpha=\kappa=1, \beta=2, r=0.05, K = 0.15, a=b=0.07$. For this set of parameters, the figure
shows the convexity of the American call price with respect to $y$.

\par} \vs{10pt}

\end{figure}

\begin{proof}

$(A)$ First, we clearly have that the conditions for the local time-space formula on curves (Peskir (2005a)) hold (in the relaxed form) for $e^{-rs}C^A(t\p s,Y^y_s)$ so that
\begin{align} \label{mix-th-3} \hs{1pc}
e^{-rs}C^A&(t\p s,Y^y_s)\\
=\;&C^A(t,y)+M_s\nonumber\\
 &+ \int_0^{s}
e^{-ru}\left(C^A_t \p\L_Y C^A
\m rC^A\right)(t\p u,Y^y_u)
 I(X^x_u \neq \{b_*(t\p u),b^*(t\p u)\})du\nonumber\\
 &+\frac{1}{2}\int_0^{s}
e^{-ru}\left(C^A_y (t\p u,Y^y_u +)-C^A_y (t\p u,Y^y_u -)\right)I\big(Y^y_u=b_*(t\p u)\big)d\ell^{b_*}_u(Y^y)\nonumber\\
&+\frac{1}{2}\int_0^{s}
e^{-ru}\left(C^A_y (t\p u,Y^y_u +)-C^A_y (t\p u,Y^y_u -)\right)I\big(Y^y_u=b^*(t\p u)\big)d\ell^{b^*}_u(Y^y)\nonumber\\
  =\;&C^A(t,y)+M_s  \nonumber\\
  &+\int_0^{s}
e^{-ru}\left(\L_Y G\m rG\right)(t\p u,Y^y_u)I(Y^y_u \le b_*(t\p u)\,\text{or}\,Y^y_u \ge b^*(t\p u))du\nonumber\\
  =\;&C^A(t,y)+M_s +\int_0^{s}
e^{-ru}H(Y^y_u)I(Y^y_u \le b_*(t\p u)\,\text{or}\,Y^y_u \ge b^*(t\p u))du\nonumber
  \end{align}
where we used \eqref{mix-PDE} and the smooth-fit conditions \eqref{mix-SP1}-\eqref{mix-SP2}, \eqref{mix-FBP2} and where $M=(M_s)_{s\ge 0}$ is the martingale term,  $(\ell^{b}_t(X^x))_{t\ge 0}$ is the local time process of $X^x$ at the boundaries $b\in\{b_*,b^*\}$
\begin{align} \label{mix-Tanaka-3} \hs{3pc}
 \ell^{b}_t (X^x):=\QQ-\lim_{\eps \downarrow 0}\frac{1}{2\eps}\int_0^{t} I(b(t\p u)\m\eps<X^x_u<b(t\p u)\p\eps)d\left \langle X,X \right \rangle_u.
 \end{align}
 Now,
upon letting $s=T\m t$, taking the expectation $\EE$, recalling the definition of $C^E$ in \eqref{mix-eur-1}, using the optional sampling theorem for $M$, rearranging terms and noting that
$C^A(T,y)=G(y)=(f(y)\m K)^+$ for all $y>0$, we get \eqref{mix-th-1}.
The system of integral equations \eqref{mix-th-2}-\eqref{mix-th-3} is obtained by substituting $x=b_*(t)$ and $x=b^*(t)$ into \eqref{mix-th-1} and using \eqref{mix-IS1} and \eqref{mix-IS2}, respectively.
\vs{6pt}

$(B)$ The proof of that the pair $(b_*, b^*)$  is the unique solution to the system \eqref{mix-th-2}-\eqref{mix-th-3} in the class of continuous functions $t\mapsto b_*(t)$ and
$t\mapsto b^*(t)$ follows from arguments similar to those employed in Theorem 3.1 in Section 3.

\end{proof}

\begin{remark} The results of this section might be seen as generalizations of the results in Sections 2-4 if we slightly change the model and take $f=f_1 +f_2$
where $f_1$ is of $(A1)$-type or zero function, and $f_2$ is of $(A2)$-type or zero. Then if $f_1\equiv 0$ (thus $f$ is of 1/2-type), we have $K_*=0$, $b_*=0$ and a single boundary $b^*$ for $Y$, which
can be translated into the boundary $f(b^*)$ for the VIX process $X$. If now $f_2\equiv 0$ (i.e. $f$ is of 3/2-type), we have $K^*=\infty$, $b^*=\infty$ and a single boundary $b_*$ for $Y$, which
can be transformed into the boundary $f(b_*)$ for $X$.
\end{remark}

\section*{Appendix}
\renewcommand{\theequation}{A-\arabic{equation}}

Here, we show that the models in Examples 2.1-2.6 satisfy
\textbf{Assumption R} under some conditions for parameters when needed.
\vs{6pt}

1. ($3/2$-model) When $\beta>\kappa^2$, we get
 \begin{align*}\hs{6pc}
 h\left( x\right) =x\left( \alpha \m r\right)
-\left( \beta \m\kappa ^{2}\right) x^{2}+rK
\end{align*}
with $
x^{\ast }=\frac{\alpha -r+\sqrt{\left( \alpha -r\right) ^{2}+4\left( \beta
-\kappa ^{2}\right) rK}}{2\left( \beta -\kappa ^{2}\right) }>0.
$
\vs{6pt}

2. ($1\p1/(2\nu)$-model) When $\beta>\frac{1}{2}\kappa ^{2}\left( \nu
\p 1\right)$, we obtain that
 \begin{align*}\hs{6pc}
 h\left( x\right) =\nu \left( \left( \alpha \m
\tfrac{r}{\nu }\right) x-\left( \beta \m\tfrac{1}{2}\kappa ^{2}\left( \nu
\p 1\right) \right) x^{1+1/\nu }\right) +rK
\end{align*}
 is a strictly
concave function for $x>0$ with $h\left( +\infty \right) =-\infty $. The
threshold $x^{\ast }$ is the unique positive root of $\nu \left( \left(
\alpha \m \frac{r}{\nu }\right) x-\left( \beta \m\frac{1}{2}\kappa ^{2}\left(
\nu \p1\right) \right) x^{1+1/\nu }\right) +rK=0$.
\vs{6pt}

3. (mixture $1\p1/(2\nu _{j})$, $j=1,...,n$ model) We were not able to verify analytically
the \textbf{Assumption R} for this model, however numerical results strongly support the claim
that this assumption is satisfied when $
\beta >\frac{1}{2}\kappa ^{2}\left( \nu_j \p1\right)$ for any $j=1,...,n$.
\vs{6pt}

4. ($1/2$-model) We have that
 \begin{align*}\hs{6pc}
 h\left( x\right) =\beta -\alpha x-r\left( x\m K\right)
\end{align*}
with $x^{\ast }=\left( \beta \p rK\right) /\left( \alpha \p r\right) $.
\vs{6pt}

5. ($1\m 1/\left( 2\nu \right) $-model) When  $%
\beta +\frac{1}{2}\kappa ^{2}\left( \nu -1\right) >0$ (which is satisfied under Feller condition $\beta\ge \kappa^2/2$ we imposed throughout the paper) we have that
\begin{align*}\hs{3pc}
h\left( x\right) =\nu x^{1-1/\nu }\left( \beta +\tfrac{1}{2}\kappa
^{2}\left( \nu -1\right) \right) -\left( r\p\nu \alpha \right) x+rK
\end{align*}
is a strictly decreasing function for $x>0$ with $h\left(+ \infty \right)
=-\infty $. The threshold $x^{\ast }$ is the unique positive root of $\nu
x^{1-1/\nu }\left( \beta +\frac{1}{2}\kappa ^{2}\left( \nu -1\right) \right)
-\left( r+\nu \alpha \right) x+rK=0$.
\vs{6pt}

6. (mixture $1-1/\left( 2\nu _{j}\right) $, $j=1,...,n$ model)
When $
\beta +\frac{1}{2}\kappa ^{2}\left( \nu_j -1\right) >0$ for any $j=1,...,n$ (which also holds under Feller condition), we have that%
\begin{align*}\hs{3pc}
h\left( x\right) =\sum_{j}\omega _{j}\nu _{j}g^{\nu _{j}-1}(x)\left(\beta+\tfrac{1}{2}\kappa^2\left(\nu_j -1\right) \right)
-\alpha \sum_{j}\omega _{j}\nu _{j}g^{\nu _{j}}(x)-r(x\m K)
\end{align*}
is a strictly decreasing function for $x>0$ with $h\left( +\infty \right)
=-\infty $. The threshold $x^{\ast }$ is the unique positive root of $\sum_{j}\omega _{j}\nu _{j}g^{\nu _{j}-1}(x)\left(\beta+\frac{1}{2}\kappa^2\left(\nu_j -1\right) \right)
-\alpha \sum_{j}\omega _{j}\nu _{j}g^{\nu _{j}}(x)-r(x\m K)=0$.

\vs{12pt}

\begin{center}

\end{center}

\end{document}